\documentstyle[12pt,aasms4]{article}

\newcommand{\ctanb}{$^{13}$C($\alpha$, $n$)$^{16}$O~}

\newcommand{\neanb}{$^{22}$Ne($\alpha$, $n$)$^{25}$Mg~}

\newcommand{\linb}{$^{7}$Li($p$, $n$)$^{7}$Be~}

\newcommand{\msb}{$M_{\sun}$~}
\newcommand{\zsb}{$Z_{\sun}$~}
\newcommand{\ms}{$M_{\sun}$}

\newcommand{\ctb}{$^{13}$C~}
\newcommand{\ct}{$^{13}$C}

\newcommand{\ndb}{$^{142}$Nd~}
\newcommand{\nd}{$^{142}$Nd}
\newcommand{\spt}{$s$-process~}

\lefthead{Arlandini et al.}
\righthead{Neutron capture in low mass AGB stars}

\begin{document}

\title{Neutron capture in low mass Asymptotic Giant Branch stars: \\
cross sections and abundance signatures }

\author{Claudio Arlandini\altaffilmark{1}, Franz K{\"a}ppeler\altaffilmark{2} and Klaus Wisshak\altaffilmark{3}}
\affil{Forschungszentrum Karlsruhe, Institut f{\"u}r Kernphysik, 
          Postfach 3640, D-76021 Karlsruhe, Germany}

\author{Roberto Gallino\altaffilmark{4}} 
\affil{Dipartimento di Fisica Generale, Universit{\`a} di Torino, 
          I-10125 Torino, Italy}

\author{Maria Lugaro\altaffilmark{5}}
\affil{Department of Mathematics, Monash University, Clayton, 
          Victoria 3168, Australia}

\author{Maurizio Busso\altaffilmark{6}} 
\affil{Osservatorio Astronomico di Torino, I-10025 Torino, Italy}

\and

\author{Oscar Straniero\altaffilmark{7}}
\affil{Osservatorio Astronomico di Collurania, I-64100 Teramo, Italy}

\altaffiltext{1}{email: claudio@ik3.fzk.de}
\altaffiltext{2}{email: kaepp@ik3.fzk.de}
\altaffiltext{3}{email: wisshak@ik3.fzk.de}
\altaffiltext{4}{email: gallino@ph.unito.it}
\altaffiltext{5}{email: marial@thala.maths.monash.edu.au}
\altaffiltext{6}{email: maurizio@otoxd6.to.astro.it}
\altaffiltext{7}{email: straniero@astrte.te.astro.it}

\begin{abstract}
The recently improved information on the stellar ($n$, $\gamma$) cross 
sections of neutron-magic nuclei at $N$ = 82, and in particular of 
$^{142}$Nd, turned out to represent a sensitive test for models of $s$-process 
nucleosynthesis. While these data were found to be incompatible with the
classical approach based on an exponential distribution of neutron
exposures, they provide significantly better agreement between
the solar abundance distribution of $s$ nuclei and the predictions of models 
for low mass AGB stars.
The origin of this phenomenon is identified as being due to the
high neutron exposures at low neutron density obtained between thermal pulses when the 
$^{13}$C burns radiatively in a narrow layer of a few 10$^{-4} M_{\odot}$.
This effect is studied in some detail, and the influence of the
presently available nuclear physics data is discussed with respect
to specific further requests. In this context, particular attention 
is paid to a consistent description of $s$-process branchings
in the region of the rare earth elements. It 
is shown that - in certain cases - the nuclear data are 
sufficiently accurate that the resulting abundance uncertainties
can be completely attributed to stellar modelling. Thus, the $s$ process
becomes important for testing the role of different stellar masses
and metallicities as well as for constraining the 
assumptions for describing the low neutron density provided by the \ctb
source.

\end{abstract}

\keywords{stars: AGB -- stars: evolution -- stars: low mass -- 
nucleosynthesis}

\section{Introduction} 

In the last thirty years studies of the slow neutron capture process ($s$ process) 
have been pursued either
through nucleosynthesis computations in stellar models for the Thermally 
Pulsing Asymptotic Giant Branch (TP-AGB) phases of low and intermediate mass 
stars (\cite{ul73}; \cite{ti77}; \cite{hi88}; \cite{kap90}; \cite{sp95}; \cite{gp98}) or by 
phenomenological models, mostly by the so-called {\it classical approach}, which were developed
with the heuristic intention to provide a possibility for a "model-free" 
description. In this last case, simple analytical expressions are used for the
neutron irradiation and, to first approximation, any time dependence of the 
physical parameters is neglected (see e.g. \cite{kap89}). For a brief 
historical account of $s$-process analyses see 
Gallino, Busso, \& Lugaro (1997).

For low mass stars, the two descriptions appeared to be compatible within the respective 
uncertainties, in particular after the \ctanb
neutron source was recognized to play a major role on the AGB
(\cite{gp88}; \cite{hi88}). There, the $s$ process was 
assumed to occur in convective thermal pulses, the classical analysis was
considered to yield "effective" conditions characterizing the stellar 
scenarios (\cite{kap90}).

This situation changed after it was realized that $^{13}$C burns 
radiatively in the
time interval between two successive convective He-shell instabilities (\cite{sp95}). 
The interplay of the different thermal conditions for the $^{13}$C and
$^{22}$Ne neutron sources, both contributing to the nucleosynthesis process, is 
hardly represented by a single set of effective parameters like
those commonly used by the classical approach.  This is particularly true for the  
description of the neutron exposure, which is usually simplified by an
exponential distribution. In contrast, present stellar models show that the
distribution of neutron exposures is definitely non-exponential, and actually very 
difficult to be described analytically (Arlandini et al. 1995). 
Any attempt to describe this picture in a phenomenological way  
would require to increase the number of free parameters, in contradiction to the basic reason for 
such an approach as a model-independent guideline for stellar calculations.

Despite of the substantial differences between classical approach and the complex AGB models,
the classical analysis is, however, still reproducing the $s$ abundances for the majority of nuclei
between Sr and Bi, simply because two conditions are commonly satisfied in the mass regions
between magic neutron numbers: the respective neutron capture rates are independent of
temperature due to the $1/v$-behavior of the ($n$, $\gamma$) cross sections and are sufficiently
large to establish steady flow equilibrium.
Consequently, the resulting $s$ abundances are to good approximation inversely proportional
to the Maxwellian averaged cross sections.

Nevertheless, the differences between classical approach and detailed AGB models
become the more evident the more accurate ($n$, $\gamma$) rates of nuclei involved in
$s$-process branchings and/or with strong deviations from a $1/v$-behavior are
becoming available.
A region of the $s$-process path where this is particularly
evident is that 
involving the neutron-magic nuclei at $N$ = 82, including the $s$-only isotope \nd.
Recent accurate measurements of the stellar neutron capture cross sections 
for all stable isotopes in this area (\cite{wip98a}, 1998b; 
\cite{vp99}) revealed a number of significant discrepancies compared to the older 
data (see compilations by Bao \& K\"appeler 1987; Beer, Voss, \& Winters 1992),
resulting in a situation where the classical 
approach implies a large overproduction of \ndb for any reasonable parameter set.
This problem was already noted by Guber et al. (1997), but their relatively large cross section 
uncertainty did not allow any firm conclusions. 
At the same time the new data were found to be fully compatible with
the results of the stellar model. 

The consequences of this discrepancy will be the subject of the present paper.
We show how the classical model leads to internal inconsistencies, 
at least near neutron-magic nuclei, which seem to be the inescapable consequence 
of this simplifying approach. On the other hand, stellar models appear
increasingly successful in describing the many facets of the $s$-abundance patterns, 
regardless of their much higher complexity. An important aspect in this discussion will be
that average quantities, such as the effective values for the neutron density and temperature
deduced in the classical approach, do not easily relate to the true features of the 
stellar scenario. That holds even for the zones of the $s$-process path where both models provide
a satisfactory reproduction of the solar $s$ abundances. Also
more complex phenomenological models with an increasing number of free parameters
lack a detailed comprehension of the physical conditions at the $s$-process site.

Finally, we note that the use of the solar $s$ abundances as a constraint for 
$s$-process studies has to be made with caution, since the $s$ process is not a 
unique event but rather the result of a complex Galactic evolution mechanism.
In particular, the $s$-process distribution varies strongly for TP-AGB stars with different 
metallicity (Busso, Gallino, \& Wasserburg 1999; Travaglio et al 1999; Raiteri et al. 1999).
Therefore, the comparison with the solar distribution has to be
complemented by direct observations in different types of $s$-enriched stars
as well by the signatures carried by interstellar grains.

In \S 2 an overview of the $s$-process models is presented. In \S 3 and \S 4 
we discuss the effect of the new nuclear physics input data on the classical approach 
and stellar model 
calculations, respectively. In \S5 the relevance of the uncertainties of some aspects of the
stellar evolution calculations on the $s$-process nucleosynthesis results are discussed. 
This is followed in \S 6 by an analysis of some relevant branchings 
of the $s$-process reaction path. In \S 7, we conclude with a presentation of the $r$-process 
residuals
obtained with the two approaches and some final remarks in \S 8. 
 
\section{The $s$-process models}

Since the early works (Seeger, Fowler, \& Clayton 1965; 
Clayton \& Rassbach 1967; Clayton \& Ward 1974), the process of 
slow neutron addition in red giants has been approximated by a 
phenomenological approach, consisting of an analytical formulation 
for the distribution of $\left<\sigma \right>^iN_{s}^i$ products, where $N_s^i$ is the
fractional $s$ abundance of nucleus $i$ and $\left<\sigma\right>^i = \left< \sigma v \right>^i/v_T$ is
its Maxwellian averaged neutron capture cross section, with $v_T$ 
thermal velocity. It is known (\cite{c68}) that
such a formulation is obtained by adopting a suitable distribution 
of neutron exposures $\rho$($\tau$), where $\tau$ means the time-integrated neutron flux.

Over the years, this procedure has been particularly successful for the 
so-called {\it main} $s$-process component, accounting for $s$-nuclei 
between magic neutron numbers $N$ = 50 and 126, i.e. for mass 
numbers 88 $< A <$ 208. Traditionally, the most common
form for $\rho$($\tau$) was an exponential distribution,
$$
\rho(\tau) = \frac {G N_{\odot}^{56}} {\tau_0 } {\rm exp}(-\tau/\tau_0),
$$
that proved very effective in reproducing the solar system $\left<\sigma \right>N_{s}$ curve
with the fit of only two parameters: the fraction $G$ of the solar iron
abundance that would be required as a seed, and the mean
neutron exposure $\tau_0$. The treatment of branchings, as formulated
by Ward, Newman, \& Clayton (1976), requires three additional parameters,
namely the temperature, $T$, the neutron
density, $n_n$, and the electron density, $n_e$.

A physical justification for the choice of 
$\rho$($\tau$) seemed to appear when Ulrich (1973) showed that
an exponential distribution of exposures was the natural consequence of
repeated He-shell flashes during the AGB phase. 
The exponential distribution was shown to derive simply from the 
partial overlap of subsequent thermal pulses. With a constant exposure $\Delta \tau$ per pulse,
and a constant overlap factor $r$, after $N$ pulses the fraction of 
material having experienced an exposure $\tau$ = $N\Delta \tau$
is $\sim r^N \equiv \exp{(-\tau/\tau_0)}$, where the the mean exposure $\tau_0$ is defined as
$\tau_0 = -\Delta \tau/\ln{r}$.

The pulsed nature of the neutron source was later recognized by the analysis of the branching
at $^{85}$Kr (Ward \& Newman 1978), and taken into account in more complex phenomenological approaches
 (\cite{b91}). The classical analysis was found to be a useful way of describing the $s$ abundances, as it 
provided an apparently consistent, simple and straightforward
description of the $s$ process, suggesting it to occur in a stellar
environment with physical parameters corresponding to
those inferred from this model. 

Meanwhile, the knowledge of the last stages of evolution for AGB stars
was improved by a series of investigations originally based on the activation of the
\neanb reaction in intermediate mass stars (IMS) (e.g. \cite{ti77}; \cite{cp80}; 
\cite{bp88}), and subsequently emphasizing the importance 
of the alternative \ctanb reaction in low mass stars (LMS) (\cite{gp88};
\cite{hi88}; \cite{kap90}). The latter models finally suggested that the
main $s$-process component results from the interplay of both neutron 
sources in LMS with masses between 1.5 and 3 $M_{\odot}$. The present status of
this field of research was recently updated on the basis of revised stellar 
models, including a self-consistent mixing mechanism for the transport of
freshly synthesized material from the He shell to the stellar surface, the
so-called third dredge-up (TDU) (\cite{sp97}). Accordingly, the abundance distributions considered in 
the following refer to the composition of the TDU material integrated
over the whole AGB phase and lost by stellar winds. This composition represents the $s$-process enrichment
of the interstellar medium by the considered model star. This has to be considered 
in comparisons with the $s$-process enhancements observed in chemically peculiar red giants
(\cite{sl85},1986,1990; \cite{lp95}; \cite{bp92},1995,1999).

According to the above LMS models, the $^{13}$C neutron source is activated under
radiative conditions during the intervals between subsequent He-shell burning episodes.
While the $^{13}$C reaction provides the bulk of the neutron exposure already at 
low temperatures ($kT$ $\simeq$ 8 keV) and neutron densities ($n_n \le$ 10$^7$ cm$^{-3}$), 
the produced abundances are modified by the marginal activation of the $^{22}$Ne source
during the next convective instability, when high peak neutron densities of $n_n$ $\le$ 
10$^{10}$ cm$^{-3}$ are achieved at $kT$ $\simeq$ 23 keV. Though this second neutron burst 
represents only a few percent of the total exposure, it suffices to modify the abundance
patterns of several temperature- and neutron-density-dependent branchings.
The time dependence of this second burst is particularly important for defining
the freeze-out conditions for most of these branchings. 

Remarkably similar physical conditions are 
found in AGB models down to a metallicity slightly lower than 1/2 solar (-0.4 
$\le$ [Fe/H] $\le$ 0) by Gallino et al. (1998). Though observations in MS and S stars in 
the solar neighborhood (Smith \& Lambert 1990; Busso et al. 1992) exhibit a spread in the respective 
$s$ abundances, most Galactic disk AGB stars in
the mass range 1.5 $\le M/M_{\odot}$ $\le$ 3 can be considered as suitable sites for 
reproducing the main component: while the solar $s$ composition is clearly the product of an average
over Galactic astration processes from various 
generations of AGB stars with different $s$-process efficiencies, according to
the initial mass, metallicity, and mass loss mechanism, it is remarkable 
that the solar $s$-abundance distribution lays roughly at 
the center of the spread observed in MS and S stars
(Busso et al. 1999). The actual neutron capture
nucleosynthesis efficiency in each star depends on the metallicity, the choice 
of the amount of $^{13}$C that is burnt, and its profile in the 
intershell region, i.e. on what has become known as 
{\it the $^{13}$C pocket}. In order to model this essential feature, which 
is controlled by partial proton mixing below the 
formal convective envelope border, a detailed hydrodynamical 
treatment of the H/He interface is required. Presently available 
hydrostatic stellar models cannot account for this phenomenon. 
Thus, in our computations it is still parameterized in a relatively 
free way (see e.g. \cite{gp98}; \cite{bgw99} for a discussion). Hence, the AGB 
nucleosynthesis results presented here, which are shown to account for 
the main $s$-process component, are based on suitable choices for the 
$^{13}$C pocket and metallicity. They should be considered
as the nucleosynthesis pattern of a particular AGB star of the Galactic 
disk, whose $s$-process abundance distribution closely matches 
the main $s$-process component. As said, it is a relatively common occurrence
in the Galaxy.

\section{The $^{142}$Nd cross section and the limits of the classical 
approach}
\subsection{New cross sections}
The $s$-process abundance of $^{142}$Nd, shielded against a $r$-process
contribution by $^{142}$Ce (Fig. 1), is influenced by two small branchings in the neutron capture 
path at $^{141}$Ce and $^{142}$Pr. Provided the $p$-process
contribution be negligible, the branching probabilities follow from the comparison
of the empirical product $\left< \sigma \right> N_s$ of the stellar
neutron capture cross section and the solar \ndb abundance with the
$\left< \sigma \right> N_s$  systematics in the local mass region.
The expected branching factor is about 5\%, so a meaningful analysis
was hampered by the uncertainties in the nuclear physics data, 
especially by the 9\% uncertainty of the \ndb cross section (\cite{bvw92}).
Recently, experimental determinations of the stellar
neutron capture cross sections of $^{140,142}$Ce (\cite{kap96}), $^{141}$Pr (\cite{vp99}) and of 
all the stable Nd isotopes (\cite{wip98a}, 1998b; \cite{tp95}) have been provided 
with uncertainties of 1-2\% along with new cross section calculations
for the unstable branch point nuclei $^{141}$Ce and 
$^{142}$Pr (\cite{kap96}). The significant discrepancies with respect to previous data
(see Table 1) triggered a detailed $s$-process analysis.
\placetable{tbl1}
\placefigure{fig1}

The stellar neutron capture cross sections of $^{142,144}$Nd were also
recently measured by Guber et al. (1997). While the values 
agree within the quoted uncertainties at 30 keV, a serious discrepancy is found for $^{144}$Nd at 
$kT$ $\leq$ 20 keV. However, this discrepancy had no effect in the present study.
The adopted Ce, Pr, and Nd cross sections were obtained at the Karlsruhe 3.75 MV Van de
Graaff accelerator. The cerium cross sections
were measured with an activation method, while the neodymium and praseodymium 
experiments were performed with the Karlsruhe 4$\pi$ Barium Fluoride 
(BaF$_2$) detector and the time-of-flight (TOF) method.

The activation method consists in irradiating a sample in a quasistellar
neutron spectrum, obtained by bombarding a thick metallic Li target with
protons of 1912 keV, just above the reaction threshold. The \linb reaction
then yields a continuous energy distribution with a high energy cutoff at
$E_n$=106 keV. The resulting neutrons are emitted in a forward cone of
120$^\circ$ opening angle. The angle integrated spectrum closely resembles
a Maxwellian distribution peaked at 25 keV, thus exhibiting almost exactly
the shape required to determine directly the stellar cross section. 
The samples are placed on the lithium target, 
sandwiched between gold foils. The simultaneous activation of the gold foils
serves for normalization, since both the stellar neutron capture cross section
of $^{197}$Au($n$, $\gamma)^{198}$Au (\cite{rk88}) and the decay parameters
of $^{198}$Au (\cite{au83}) are accurately known.
A more detailed description of the method and of the experimental setup can be found
in Beer \& K\"appeler (1980).
After activation, the $\gamma$-rays from the decay of the product nuclei
are counted with a high-purity Ge-detector.

As for the TOF experiments, the neutron energies
were determined by time of flight, with the samples being located at a flight path of 
79 cm. Adjusting the proton energy slightly above the reaction threshold a continuous 
neutron spectrum in the energy range relevant for the determination of the Maxwellian 
averaged stellar cross sections, i.e. from 3 to 200 keV, is obtained. 
Capture events were registered with the Karlsruhe
4$\pi$ Barium Fluoride detector via the prompt capture $\gamma$-ray cascades.
This detector consists of 42 hexagonal and pentagonal crystals forming a spherical 
shell with 10 cm inner radius and 15 cm thickness. It is characterized by a 
resolution in $\gamma$-ray energy of 7\% at 2.5 MeV, a time resolution of 500 ps,
and a peak efficiency of 9\% at 1 MeV. A comprehensive description can be found
in Wisshak et al. (1990).
Again, the stellar cross sections are calculated using gold as a standard.

The TOF method represents a universal approach, the only limitation being a minimum 
sample mass. On the other hand, the activation technique offers a far superior sensitivity,
since the samples can be placed directly at the neutron production target in
a much higher neutron flux. However, this method can only be applied to cases where
the product nucleus is unstable. Also, the systematic uncertainties
are somewhat larger than those typical for the TOF method.

The cross sections being determined with uncertainties at the 1-2\% level, the
$s$-process abundances are derived with similar accuracy. Therefore, 
possible $p$-process contributions can no longer be neglected. This correction is most 
important for the $s-only$ nuclei, which are shielded only against 
the $r$-process $\beta$-decays. An empirical determination
based on the abundances of local $p-only$ isotopes would suggest a contribution of
$\sim$9\% for \nd, but available $p$-process calculations (\cite{rp90}; \cite{pp90}; \cite{hp91}; 
\cite{rpc91}; \cite{hpc91}; \cite{rp95}) give lower values of $\sim$4\%. Such
comparably large $p$-process abundance is plausible,
since \ndb is the heaviest neutron-magic stable nucleus with $N$ = 82 and is, therefore,
favored in a $p$-process environment, where the ($\gamma$, $n$) flow is damped at the 
higher neutron binding energies. Additionally, \ndb is enhanced 
by the decay of its 
$\alpha$-unstable $p$-process progenitors $^{146}$Sm, $^{150}$Gd, and $^{154}$Dy.  
Another small correction refers to the effect of thermally populated excited nuclear states.
However, such possible enhancements of the ($n$, $\gamma$) cross sections do not influence
the rates used in this discussion.

\subsection{The classical approach}

\ndb is located immediately at the pronounced precipice of the
$\left< \sigma \right>N_{s}$ curve that is caused by the small ($n$, $\gamma$)
cross sections at $N$ = 82. Hence, its cross section 
determines not only the branching analysis, but also the general shape of 
the $s$-process distribution, which is described in the classical approach via 
the mean exposure, $\tau_0$. 
Since the $\beta$-decay rates of $^{141}$Ce and $^{142}$Pr are almost independent
of the stellar temperature (\cite{ty87}), these branchings will be completely defined 
by the effective $s$-process neutron density. This parameter can best be determined 
by means of the $s$-only pair $^{148}$Sm and $^{150}$Sm. As the result of
three branchings at $^{147}$Nd, and $^{147,148}$Pm, the first isotope, $^{148}$Sm,
is partially bypassed while $^{150}$Sm experiences the entire $s$-process flow.  
The {\it effective} neutron density is found to be $n_n = \left( 4.1\pm0.6 \right)\times 10^8$
 cm$^{-3}$ (\cite{tp95}). A more accurate determination requires an experimental value for
the stellar neutron capture cross section of $^{147}$Pm.

Following the concept of Ulrich (1973), classical $s$-process calculations have 
been performed by means of the network code NETZ (\cite{j90}), using a neutron 
density of $4.1 \times 10^8$ cm$^{-3}$.
The adopted thermal energy was $kT$ = 30 keV, according to the analysis
of Wisshak et al. (1995) on the branchings bypassing the $s$-only isotopes $^{152,154}$Gd,
which constrained the $s$-process temperature to 28 $\le$ $kT$ $\le$ 33 keV.
The electron density $n_e$ = $5.4 \times 10^{26}$ cm$^{-3}$ was obtained from
a reanalysis of the branching at $A$ = 163 feeding $^{164}$Er (\cite{b96}; \cite{a98}).
The best fit of the solar system distribution for the $s$-nuclei belonging
to the main component is obtained for $\tau_0$=(0.296$\pm$0.003) [$kT$/30]$^{1/2}$ mbarn$^{-1}$,
slightly lower than the previously adopted value of (0.300$\pm$0.009) [$kT$/30]$^{1/2}$ mbarn$^{-
1}$ (\cite{kap90}, correctly extrapolating the value given at $kT$ = 29 keV).
Note the significant reduction in the uncertainty of $\tau_0$, 
which is due the improved cross sections around $N$ = 82. The quality of the fit
is evaluated by using the unbranched $s$-only nuclei as normalization points, i.e.
$^{100}$Ru, $^{110}$Cd, $^{116}$Sn, $^{122,123,124}$Te, $^{150}$Sm, and $^{160}$Dy.
The root mean square deviation of their calculated $\left<\sigma \right>N_s$ values
from the respective empirical points is
$$\delta = \left[ \frac{1}{n} \sum \frac{ \left(\left<\sigma \right> N^{calc} - 
\left<\sigma \right> N^{emp}\right) ^2}{\left( \left<\sigma \right> N^{emp} \right) ^2} \right]^{1/2} 
= 6\% .$$
This mean deviation is larger than reported previously (\cite{kap90}), since it includes 
$^{160}$Dy, whose cross section has been experimentally determined and may well be influenced by
the stellar temperature (\cite{vp99}). Without $^{160}$Dy, this value would reduce to 4\%. 
\placefigure{fig2}

Fig. 2 (bottom left panel) shows that the $^{142}$Nd is overproduced by
$\sim$12\%, despite the fact that it is partly bypassed by the reaction
flow. This overproduction is neither compatible with the 2\%
cross section uncertainty at 30 keV, nor with the uncertainty of the solar abundance, 
because the 
abundance ratio between the chemically related rare earth elements Nd
and Sm is known to the 1.8\% level (\cite{ag89}). 
Furthermore, the overproduction factor has to be considered as a lower limit,
due to the non-negligible $p$-process contribution. Accordingly,
$^{142}$Nd is the first clear evidence that the simple assumptions of the 
classical model are not adequate to describe the $s$ process. 

Previously,  
similar difficulties of 
the classical model were noted already in connection with the notorious underproduction of $^{116}$Sn
and the overproduction of 
the $s$-only isotope $^{136}$Ba (\cite{vp94}; \cite{wip96}), but in these cases the solar Sn and Ba 
abundances 
were too uncertain to allow for a conclusive argument.
Another problem of the classical approach was related to the overproduction
of $^{86}$Kr and $^{87}$Rb (see e.g. \cite{kap90}) due to the branching at $^{85}$Kr. Even if the 
contribution from the $r$ process and from the weak
component are neglected, the classical approach yields large overabundances with respect to
the other $s$ nuclei. A possible solution was suggested by
the assumption of a pulsed $s$ process in more complex phenomenological approaches 
(Ward \& Newman 1978; Beer \& Macklin 1989; \cite{b91}), constraining the neutron 
pulse duration to 3 yr $\leq \Delta t$ $\leq$ 20 yr. 

\subsection{Other phenomenological models}

In order to understand if the failure of the classical model may be 
considered in a more general way,  
we analyzed also other parametrized approaches.
\cite{sp65}, following a suggestion by \cite{cp61}, found that the solar 
$s$-distribution can be adequately fitted by a discrete superposition
of a limited number (four in their example) of single neutron exposures.
We computed a grid of 50 distributions for single neutron exposures
with $\Delta \tau$ ranging from 0.03 to 3.50 mbarn$^{-1}$, assuming a constant neutron density
of $4.1 \times 10^8$ cm$^{-3}$ and 
varying the irradiation time. The solar distribution
of all $s$-only isotopes was fitted from Fe to Pb as a weighted sum of four exposures,
using a $\chi^2$ method.
The most promising results fall in two groups, the first consisting of
solutions that are very similar to those obtained
with an exponential exposure distribution for the main component. These cases provide a good
overall reproduction of the solar abundances, but yield strong
overproductions of \ndb and $^{136}$Ba. They include also a 
very small exposure, which mimics the so-called {\it weak component}, representing the
$s$ process in massive stars. This component accounts for most of the 
$s$ abundances below $A$ = 90.
The second group allowed for an excellent reproduction of all
$s$-only isotopes from Te to Sm, but led to unavoidable
and unacceptable overproductions of more than 10\% for all isotopes
lighter than Te or heavier than Sm, including the important
normalization points $^{100}$Ru and $^{110}$Cd.
Therefore, this idea of a superposition of a limited number of single
neutron exposures was found to be no more successful than the classical approach.

A model describing the pulsed
burning of two neutron sources with an exponential distribution
of exposures was suggested by Beer (1991) and Beer et al. (1997).  Although the number of free 
parameters is more than doubled with respect to the classical model, the fit to 
the solar main $s$ component is only slightly
better, since it does not solve the Sn and Ba problems, nor
is it able to provide consistent constraints on the astrophysical
conditions of the $s$ process.

The same problems are not solved by the model proposed by \cite{g97}, which
fits the solar distribution with a large grid of components, each
characterized by a given neutron irradiation and different 
constant temperatures and neutron densities. Using an iterative
inversion procedure and without setting any predefined limit to the parameter space,
the solution shows a distribution of exposures that is very close
to an exponential law. 

\section{The new $^{142}$Nd cross section: Success for the AGB model}

A thorough description of the model and the adopted 
reaction network can be found in Gallino et al. (1998). The neutron capture calculations start from
evolutionary computations for low mass stars up to end of the AGB
phase, which were made with the latest version of the FRANEC code 
(Chieffi \& Straniero 1989; \cite{sp97}; Chieffi, Limongi, \& Straniero 1998) for a range of initial 
masses, 1.5 $\leq M_{\odot} \leq$ 3, and 
metallicities, -0.4 $\leq$ [Fe/H] $\leq$ 0. In addition, the 
influence of various mass-loss rates was checked as well. No TDU was found for lower masses. 
A satisfactory reproduction of the solar distribution of 
$s$ isotopes in the range 88 $< A <$ 208 can be obtained for all these stars,
since their physical conditions are quite similar. In this way, the only free parameters
of the neutron capture model, which are the total amount of \ctb burnt
and its profile in the pocket, could be constrained. As a general rule it was found that very similar
$s$-process abundance distributions can be obtained by contemporarily
decreasing the metallicity and increasing
the amount of \ctb in the pocket  by the same factor. 
The object of this paper is to focus on the nuclei involved in branchings
along the $s$ path, and in particular on the effect of the \ndb cross section.  
\placefigure{fig3}
 
Following Gallino et al. (1998), we consider the best representation of 
the main component obtained for a star of 2 $M_\odot$, $Z=1/2$ $Z_\odot$, and a 
Reimers mass loss rate with $\eta$ = 0.75, as a standard case.
The general improvement with respect to the classical solution (Fig. 3) is striking,
especially since no fitting procedure was applied.   
$^{134,136}$Ba are now overproduced by a mere 5\%, well compatible with 
the uncertainties of the neutron capture cross sections and solar abundances,
not requiring any more the large corrections (20\%) to the solar barium advocated 
by the classical analysis (\cite{vp94}).
Also $^{116}$Sn, for which the classical analysis of Wisshak et al. (1996) suggested a 15\% 
variation to the solar tin abundance,
is now reproduced within the respective uncertainties.
In this context, it is important to note that the meteoritic abundances of Ba and Sn 
quoted by Anders \& Grevesse (1989) were
confirmed by an independent measurement (\cite{drl98}). A similar measurement of the  
Te abundance, which is yet uncertain by 10\% would be important as well. 

As already shown
by Straniero et al. (1995), the most relevant difference compared to the 
classical analysis is actually found in the 
mass region $A$ $<$ 88. Indeed, all these nuclei 
(that are predominantly due to the weak $s$-component) are produced in much smaller quantities, 
even with respect to superseded stellar models, which assumed the convective burning of
\ct. This difference is caused by the very high neutron exposures reached
in the tiny pocket, which favor the production of heavier elements. 
In particular, at the $s$-termination path, $^{208}$Pb
is produced four times more than in the classical approach. 

This has obvious consequences for the branching at $^{85}$Kr. 
The low neutron density of the $^{13}$C neutron release implies that the reaction flow to
the neutron-rich nuclei $^{86}$Kr and $^{87}$Rb is weak, thus avoiding the overproduction 
of these isotopes that is a severe problem in the phenomenological approaches and in the 
old stellar models (\cite{kap90}). In the present case the contribution from the 
weaker exposure due to the $^{22}$Ne
neutron source is small, because of the small cross section of $^{85}$Kr.

The $^{85}$Kr branching regulates also the Rb/Sr ratio, which has been
measured in AGB stars showing \spt and carbon enrichments. With the definition [$X$] 
$\equiv$ log$_{10}(X_{star})$ - log$_{10}(X_{\odot})$, the average abundance ratio was found to be
[Rb/Sr] = -0.80 $\pm$ 0.15 (Lambert et al. 1995, Lambert 1995). Again, the new stellar model 
calculations are in good agreement with this stringent constraint.

As for the other significant differences compared to the classical approach, 
the analysis of the branchings to $^{170}$Yb and $^{192}$Pt is hampered 
by the poor knowledge of the relevant stellar ($n$, $\gamma$)
cross sections, while the branchings to $^{180}$Ta  (\cite{sp98}) and $^{176}$Lu 
(\cite{dp99}) remain
uncertain because of the complex effect of the stellar temperature on the
population of the respective ground and isomeric states.
 
In the region around $A \sim$ 140, two aspects are immediately evident from Fig. 2.
First, the new cross sections improve the situation in the stellar model, 
from a $\sim$30\% underproduction of $^{142}$Nd to 4\%, well compatible with the 
predicted $p$-process abundance.  
On the other hand, the classical approach is facing the inherent overproduction 
of $^{142}$Nd discussed in \S 3.2. The reason for reproducing $^{142}$Nd correctly 
lies in the fact that the cross section deviates significantly from
a 1/v-behavior. Accordingly, the $^{13}$C source produces 8\% less $^{142}$Nd than
the classical approach, which operates at $kT$ = 30 keV.  
In the subsequent burst from the $^{22}$Ne source relatively high peak 
neutron densities are reached for about one year, followed by a rapid freeze-out
(Fig. 4, top panel). Due to its rather small cross section, the abundance 
of \ndb itself is depleted
by no more than 10\% during this phase, the initial level being
rapidly restored during the decline of the neutron density.  
\placefigure{fig4}

Secondly, the revised \ndb and $^{144}$Nd cross sections affect the distribution 
of $s$ abundances up to $A \sim$160 by a ``propagation effect''. Indeed,
during the \ctb phase the neutron exposure is large enough to establish reaction 
equilibrium, producing an abundance reservoir at the neutron-magic 
nuclei. While this equilibrium is practically maintained
during the peak neutron density of the subsequent $^{22}$Ne burst,
the decline of the neutron density leads to pronounced freeze-out effects
near neutron-magic nuclei. 
Evidently, the abundances
of the nuclei with the smallest cross sections
freeze-out at first, whereas the isotopes beyond $^{144}$Nd are depleted 
by further neutron captures. This effect causes an additional abundance difference 
between $^{142}$Nd and $^{150}$Sm. 

A further increase of this abundance difference results from the fact that the
He shell is enriched in $s$-process material during the AGB phase
due to the overlap of subsequent He-shell flashes. 

The combination of all three effects accounts for the 12\% discrepancy
in the $^{142}$Nd abundance between the classical approach and the stellar solution.  

\section{STELLAR EVOLUTION ASPECTS}

\subsection{Parameters of the stellar models}

The presented results are affected by two kinds of uncertainties, 
those connected with the neutron capture process itself and those related to the stellar
evolutionary calculations. 

With respect to the parameters
of the stellar model, it was already emphasized that the amount of primary \ctb and its profile in the 
pocket
cannot be obtained through canonical evolutionary models and have to be, therefore, 
parameterized.  The reasons for the choice of the adopted 
profile are described by Gallino et al. (1998).

With the evolution of temperature and density in the \ctb pocket provided by the 
evolutionary models, the neutron release is determined by the rate of the 
\ctanb reaction. At low burning temperatures around  $kT$ = 8 keV, this rate is rather
uncertain since it is extrapolated 
from experimental data at higher energies (\cite{dp95}). Would the
rate be substantially lower, part of the \ctb nuclei would remain unburned and 
engulfed by the successive thermal pulse. In this case, the remaining \ctb
would be burnt at higher temperature, substantially affecting the final abundance pattern.

Therefore, test calculations have been performed by reducing the rate of Denker et al. (1995)
by factors of 2 and 10. The results show that also
for the extreme case all the \ctb would be burnt radiatively due to the progressive
increase of temperature and density in the pocket just prior
to the thermal instability. Of course, the time evolution and the peak
value of the neutron density would be different, but these variations would cause no
significant effect on the results. 

The rate of the \neanb reaction exhibits large 
uncertainties at $s$-process temperatures due to the possible existence of a 
low-lying resonance, which could substantially enhance the rate (\cite{kap94}).
Accordingly, a series of calculations was performed for the model star, varying 
the rate from the lower limit, where the possible
contribution of the 633 keV resonance was excluded, to the upper limit, which
included this resonance in toto. 

In all calculations, the standard deviation for the set of unbranched normalization 
isotopes remains less than 3\%, except for the most extreme case. As far as the 
branchings are concerned, it has to be stressed that the clear distinction of 
temperature and neutron density effects - which is made in the classical approach
- makes no sense for the stellar model, where the time-dependent
temperature profile is inherently provided by the evolutionary code.
Accordingly, it depends only on the stellar mass and metallicity and varies with
pulse number. Therefore, an increase in peak temperature implies a corresponding increase
in the peak neutron density, regardless of adopted \neanb rate. 

While the \neanb rate has no significant impact for the unbranched normalization
isotopes, it governs the abundance patterns in the $s$-process branchings.
It was found that the branchings represented by the isotope pairs 
$^{87}$Rb/$^{87}$Sr, $^{96}$Zr/$^{96}$Mo, $^{134}$Ba/$^{136}$Ba, $^{152}$Gd/$^{154}$Gd, and
$^{176}$Lu/$^{176}$Hf are most sensitive to the \neanb rate. 
Overall, the \neanb burst modifies the distribution produced by the first neutron source, 
but this remains a ''local'' process that does not reach beyond the magic barriers.
Highly non solar patterns were obtained for all isotopes listed above using rates including 
more than 50\% of the hypothetical 633 keV 
resonance.  These cases yield also unacceptable values for 
somewhat less sensitive branchings, like those bypassing $^{170}$Yb and $^{192}$Pt.
As for $^{142}$Nd, an increasing \neanb rate leads to a depletion of this nucleus 
in the neutron density maximum, resulting in a progressive decrease of 
the \nd/$^{150}$Sm ratio.

The best reproduction of the solar branching patterns is obtained with the 
recommended rate of K\"appeler et al. (1994), after excluding the hypothetical contribution 
of the 633 kev resonance. Therefore, this rate was considered as a 
''standard'' choice, although any value between the lower limit to the recommended 
value of K\"appeler et al. (1994) still provides satisfactory results. Of course, the intention of 
this analysis is not to constrain the \neanb rate, since the fine-tuning is 
dependent on the model star, but to study the sensitivity of the results with 
respect to this uncertain rate.  

With the standard choice for the \neanb rate, the \ctanb neutron burst was investigated 
by varying the quantity of \ctb in the pocket. The best fit to the solar
$s$ abundances was obtained by Gallino et al. (1998) with an average \ctb mass fraction  
of 6 $\times$ 10$^{-3}$. A series of test calculations was performed, 
by keeping the same slope of the \ctb pocket as in Gallino et al. (1998) but varying
within a factor 1.5 up and down the total amount of \ctb nuclei present in the pocket
(and using the standard stellar model for 2 $M_\odot$, 
$Z=1/2$ $Z_\odot$, $\eta$ = 0.75), a range that relates to a reasonable representation
of the main component.

As expected, the $s$-process yields in the investigated range are correlated with the amount 
of \ctb in the pocket, changing from 40\% to 170\% compared to the best representation.
Within these limits the $s$-abundance distribution is fairly well reproduced over the entire mass range of
the main component, except for the extreme cases. Larger variations than in the present test,
however, lead to important deviations. 
Moreover, it turned out that the precise reproduction of $^{134,136}$Ba 
abundances constitutes a more stringent constraint in the above test. Indeed, both nuclei are easily 
severely
overproduced for low and high values of \ct. Accordingly, this reduces the acceptable 
\ctb values to a range of $\pm$10\% around the best fit case. 
A similar lower limit is obtained from the significant
overproduction of all isotopes below $A$ $\sim$ 120.

\subsection{Uncertainties due to the evolutionary models}

The most crucial points are the mass fractions $\Delta m$ dredged up after each thermal
instability and the mass-loss rate. 
Both problems are somehow related, because the number of thermal instabilities
with TDU is determined by the mass loss rate. According to Straniero et al. (1997) the 
TDU ceases when the envelope mass becomes smaller than about 0.5 \ms.
The evolutionary code FRANEC finds TDU for stellar masses above 1.5 $M_\odot$ at
solar metallicity. The efficiency of the phenomenon is still a very debated 
matter (\cite{fl96},1998). In the present context, we consider only the uncertainties 
related to the FRANEC code.

The determination of $\Delta m$ is the most difficult problem for stellar $s$-process models,
which has a strong impact on Galactic chemical evolution. At present, the concepts for 
describing the H-He interface, a very thin zone 
compared to the mass of the envelope, exhibit a number of persisting
uncertainties. The calculated values for $\Delta m$ appear plausible
due to the constraints from stellar observations (\cite{bgw99}), but it is certainly
difficult to derive the related uncertainties from first principles.

Another problem affecting the final $s$-process abundance distribution in the envelope is related
to the uncertainty of the choice of the mass loss rate. However, an asymptotic $s$-process
distribution is reached after a limited number of pulses, so that mass loss uncertainties affect mainly
the total yield of $s$-processed material, and not much the shape of the distribution. 
This last, however, is sensitive to uncertainties in contributions from the small neutron exposure released
by the \neanb source. Since the maximum bottom temperature increases from pulse to pulse, affecting
the strength of the neutron burst, the cumulative $s$-process distribution in the envelope 
can in fact be influenced by whether the combined effects of recurrent TDU episodes and 
mass loss allow the material from the very last pulses to contribute or not.
The TDU efficiency rises 
rapidly to an almost constant value until it drops when the envelope mass
becomes sufficiently low. The effect of a larger mass loss rate was, therefore, studied by 
omitting the three last pulses, which have the highest temperatures at the bottom 
of the He burning zone. The resulting effect on the $s$-process abundances was
negligible. The only noticeable difference of about 8\% was found for $^{96}$Zr,
which is very sensitive to the neutron density.

\subsection{Influence of the initial stellar mass}

Though the nucleosynthesis yields of the investigated stars from 1.5 to 3 \msb
span a factor of two, the respective abundance distributions  
are rather similar, an important result with respect to Galactic evolution.
Moderate differences in the $s$ abundances are due to the higher 
temperatures reached during the thermal pulses in more massive stars 
(e.g. in the 3 \msb model). This implies a stronger influence
of the $^{22}$Ne source, which affects the contribution of the second burst 
to the total neutron exposure as well as the abundance patterns of several branchings.

The effect on the branchings is of the order of 5\%, except for $^{96}$Zr, which increases by a 
factor two, reaching 80\% of the average overabundance of the $s$-only isotopes in the 3 \msb star. 

\section{Relevant branchings in the $s$-process path}

\subsection{Nd-Pm-Sm}

The abundance of $^{148}$Sm is determined by the branchings at $^{147}$Nd and 
$^{147,148}$Pm, while the short lifetimes of $^{148}$Nd and $^{149}$Pm leave the 
second $s$-only samarium isotope, $^{150}$Sm, virtually unbranched.
For the involved Nd and Pm branching points, the beta-decay rates are almost independent 
of $T$ and $n_e$. Although experimental data for the cross 
sections of the unstable Pm isotopes are not yet available, the accurate measurements 
for Nd (\cite{tp95}) and Sm (\cite{wip93}) isotopes and the accurate solar abundances 
of these elements (1.3\%, \cite{ag89}) allow the most constraining 
determination of the effective neutron density $n_n$ = 4.1 $\pm$ 0.6 $\times$ 
10$^{8}$ cm$^{-3}$ (\cite{tp95}) via the classical approach. This result is
indicated in Fig. 4 (top panel) by the shaded band.

In the stellar model, the mild neutron densities of the \ctb source 
are not sufficient for the reaction flow to bypass $^{148}$Sm, thus producing it abundantly. The
opposite situation prevails, when the neutron density in the second burst 
reaches up to 10$^{10}$ cm$^{-3}$. Then, $^{148}$Sm is almost completely bypassed,
leading to a strong depletion (Fig. 4, middle panel). However, during the decline of the neutron
density, the branchings to $^{148}$Sm are restored. Eventually, the final value is established during 
the freeze-out of the abundance pattern.
Thus, in the stellar model this branching depends on the neutron density in 
a two-fold way: from the peak neutron density, which causes the 
initial destruction and explains why different stellar masses produce small but 
noticeably different results, and - predominantly - from the freeze-out of the neutron supply, 
which determines the final $^{148}$Sm/$^{150}$Sm ratio. 

It was pointed out by Cosner, 
Iben, \& Truran (1980) and K\"appeler et al. (1982) that the effective parameters 
obtained by the classical analysis have to be considered as local features, 
which, therefore, could be compared to the freeze-out conditions obtained by the 
stellar models. Reasonable agreement was found for the superseded stellar models when 
\ctb was assumed to burn convectively (K\"appeler et al. 1990).
The most intuitive criterion for the determination of the freeze-out conditions is 
to consider the moment in which the isotopic abundances reach the 90\% of their 
final values. The more complex criterion proposed by Cosner, 
Iben, \& Truran (1980) was also considered, but with negligible differences.
According to Fig. 4, the neutron density at freeze-out obtained with the present stellar 
model  is $\approx 1 \times 10^{8}$ cm$^{-3}$, considerably lower than the
phenomenological estimates. This emphasizes that the effective parameters of the classical
approach are certainly not adequate to describe the dynamical $s$-process conditions
during the AGB phase.

\subsection{Sm-Eu-Gd}

The abundance of the $s$-only isotope $^{152}$Gd is determined by branchings at $^{151}$Sm and 
$^{152}$Eu, while $^{154}$Gd is partly bypassed due to a branching at $^{154}$Eu. The $\beta$-decay
rates of $^{151}$Sm and $^{154}$Eu are extremely sensitive to the temperature, 
in particular that of $^{151}$Sm, which is enhanced by bound state decays (\cite{ty87}).
In both cases, the effect of the electron density on the stellar decay rates 
has also to be considered. 

Since both the $s$-only isotopes are partially bypassed, the reaction flow is
normalized at $^{150}$Sm. This introduces an uncertainty of only 1.3\%, since the 
relative elemental abundances are well defined in the region of the rare earth 
elements (\cite{ag89}). The main difficulty in determining the effective 
$s$-process temperature from these branchings is due to the rather uncertain 
$p$-process contribution to $^{152}$Gd. The empirical extrapolation from
neighboring $p$-only nuclei suggests this contribution to reach $\sim$30\%, whereas 
the most recent model calculations yield only a value of $\sim$12\% (\cite{rp90}; \cite{pp90}; \cite{hp91}; 
\cite{rpc91}; \cite{hpc91}; \cite{rp95}). Furthermore, a contribution of 
$\sim$6\% to $^{152}$Gd is expected from the $s$ process in massive stars (\cite{rp93}),
even if these calculations need to be updated. 

The branchings to $^{152}$Gd and $^{154}$Gd differ significantly. While $\sim$90\% 
of the flow is bypassing $^{152}$Gd, which means that $f_\beta$ is dominated 
by the $\beta$-decay rates rather than by the neutron capture rates, the branching to
$^{154}$Gd exhibits the opposite behavior. The neutron capture cross sections of 
all stable nuclei have been recently remeasured with considerably improved accuracy
(\cite{wip93}; \cite{wip95}; \cite{jp99}). New calculations have been performed 
for the branch-point nuclei (\cite{tp95}; \cite{jp99}) including an evaluation of the 
stellar enhancement factors (\cite{jp99}). For the Eu isotopes, these results are found to be 
significantly different as compared to the earlier calculations of \cite{hp76} and \cite{h81}.

Despite of the remaining uncertainties, the range of possible $s$-process temperatures
was constrained by the classical analysis, corresponding to thermal energies between 
$kT$ = 28 and 33 keV (\cite{wip93}).

On the contrary to what happens in the classical scenario, the situation in the 
stellar model is rather complex. During the \ct -burning phase, the reaction flow is almost totally 
passing through $^{152}$Gd, because of the low neutron density and since the lifetime of 
$^{151}$Sm is strongly reduced already at $\sim$8 keV. Accordingly, in this phase $^{154}$Gd 
results mainly from neutron captures on $^{152}$Gd and $^{153}$Gd, leading to a constant 
$^{152}$Gd/$^{154}$Gd ratio. Therefore, it is only during the second neutron burst that the 
thermometer-like property of these branchings becomes apparent. Due to the high peak neutron density the
reaction flow essentially bypasses $^{152}$Gd, which is almost totally destroyed, and not efficiently 
restored 
during freeze-out so that the final abundance remains relatively small. 

$^{154}$Gd is less depleted during the peak neutron density of the second burst,
reaching about 10\% of its abundance prior to the thermal instability (Fig. 4, bottom panel). 
This is so because the main reaction flow shifts only from the Gd to the Eu isotopes. 
During freeze-out, this shift is reversed, leading to the relatively high value of
the final $^{154}$Gd abundance.   

\subsection{Other branchings}

Apart from the two examples discussed in detail, there are a number of other important 
branchings along the $s$-process path, i.e. those to the $s$-only isotopes $^{164}$Er,
$^{176}$Lu/$^{176}$Hf, $^{186}$Os, and $^{192}$Pt. Except for the very complicated case at $A$ = 176,
the respective abundance patterns are well reproduced by the stellar model as can be 
seen from the regular overproduction factors of the related $s$-only nuclei (Fig. 3).

Therefore, it can be concluded, that the very sensitive test via the $s$-process
branchings has led to another confirmation of the stellar model.

\section{$r$-Residuals}

The success of the stellar model in reproducing the $s$-process pattern 
of the heavy elements opens the possibility for decomposing the solar 
abundance distribution into the respective $s$- and $r$-process components.
As far as the $p$ process is concerned, even the refined $s$-process analyses
based on accurate cross sections are not yet reliable enough to obtain quantitative
estimates for the much smaller $p$-process yields. 
\placetable{tbl2}
\placefigure{fig5}

Although the decomposition into $s$- and $r$-process components requires a 
full calculation for the $s$ abundances integrated over the Galactic evolution, 
the present stellar model results appear already to be a reasonable 
representation of the $s$-process part (Fig. 3). Therefore, the 
residuals $N_r$ = $N_{\odot}$ - $N_s$ were calculated using the $s$ abundances
obtained via the classical approach and as the arithmetic average of the 
1.5 and 3 M$_\odot$ models at $Z$ = 1/2 \zsb best reproducing the main component.

The results are listed in Table 2 for the adopted stellar model (columns 3 to 6)
and for the classical approach (columns 7 to 10). Both calculated $s$ distributions were
normalized using the solar abundance of $^{150}$Sm as a reference. Accordingly, the $r$
residuals are expected to reflect the solar $r$ distribution. The uncertainties $\delta N_s$
and $\delta N_r$ are determined by the uncertainties of 
the respective cross sections and solar abundances.    
Since stellar spectroscopy often yields elemental abundances the corresponding values are included 
in Table 2 for each element by summation over the isotopic data.
Relative overabundances in the calculated distributions with respect 
to solar are indicated by boldface numbers.

In the mass region of the main component, i.e. between Sr and Tl, 
the comparison of both distributions in Fig. 5 
shows pretty good agreement. This consistency in the $r$ residuals is reached since 
the uncertain abundances of some $s$-only isotopes are replaced by the abundances 
of their $r$-only isobars. Note, that the $r$ residuals for
Pb, and Bi, are omitted because these isotopes are significantly produced in low metallicity
stars (see Gallino et al. 1998).

In case of the classical model, the $r$ residuals have been complemented
below $^{88}$Sr by considering the parameterized weak component of Beer, Walter, \&
K\"appeler (1992). Though this schematic approach does not relate to any
realistic model, it accounts for the abundances of the respective $s$-only nuclei and may be useful
for comparison with $r$-process calculations. 

In summary, the $r$ residuals constitute a fairly robust distribution, which 
can well be used for comparison with $r$-process model calculations or astronomical
observations.

\section{Summary and conclusions}

The $s$ abundances for the main component in the mass region 88 $<$ $A$ $<$ 208
were investigated with updated ($n$, $\gamma$) rates by means of the classical 
approach and with refined stellar models for AGB stars in the range
between 1.5 and 3 \ms. Both models were found to reproduce the ensemble of those $s$-only 
isotopes, which are not affected by branchings in the reaction path, i.e. $^{100}$Ru, 
$^{110}$Cd, $^{116}$Sn, $^{122,123,124}$Te, $^{150}$Sm, and $^{160}$Dy, 
with a mean deviation of a few percent.

However, striking discrepancies between the two models were found for a few isotopes.
The most significant of these refers to $^{142}$Nd. The abundance of \ndb is affected by 
branchings in the neutron capture path at $^{141}$Ce and $^{142}$Pr, which are almost 
independent of the $s$-process temperature and electron 
density (\cite{ty87}). These branchings, and consequently the $s$ abundance of $^{142}$Nd
could be reliably characterized by means of a complete set of recently reported, 
accurate ($n$, $\gamma$) cross
sections. The significant revision of the $^{142}$Nd cross section eliminated the problem
of a persisting underproduction of this isotope by the stellar models (\cite{gp98}). In turn,
the new data imply that the classical model is now producing an inherent overabundance 
of \ndb with respect to the average of the other $s$-only nuclei, exceeding 
the respective 1$\sigma$-uncertainties by a factor 6. 
Similar but less stringent discrepancies were also found earlier 
for $^{136}$Ba (\cite{vp94}) and $^{116}$Sn (\cite{wip96}). This must be considered as evidence that 
the static assumptions for the $s$-process site, which are implicit for the 
classical model, are not realistic. The same argument applies to other phenomenological models. 

The stellar models based on recent evolutionary calculations of low mass AGB stars
are found increasingly successful in reproducing the solar distribution of $s$-nuclei.
In the light of the improved cross sections, these models were found to reproduce the 
observed $s$ abundances within the respective cross section and/or abundance uncertainties,
despite the complex scenario 
and a number of remaining problems. Another example along these lines are the large abundances in the 
Kr-Rb-Sr region predicted by the classical approach, 
which are incompatible with the additional contributions from 
the weak component and from the $r$ process. This problem does not exist in the stellar model, where the 
$s$-process production is much less efficient in this mass region, in full agreement with the low Rb/Sr 
ratios from spectroscopic observations (\cite{lp95}). This success not only refers to the overall $s$ 
distribution but is also confirmed by the proper reproduction of the abundance pattern of
the branchings in the reaction path, which represent a sensitive test for any $s$-process model.
So far, only a few branchings are determined with sufficient accuracy so that
they can be used to derive sufficiently stringent constraints. Among these are 
the branchings in the region of the REE, which all have well defined 
abundances. In particular, this has been demonstrated by the branchings characterized by the $s$-nuclei of 
neodymium, samarium, and gadolinium.  

In terms of the chemical evolution of the Galaxy the analysis of the 
$s$-process yields from AGB stars of different mass and metallicity (\cite{tp99}; Raiteri et al. 1999) confirms that the 
elements 
heavier than Ba 
including the large $^{208}$Pb abundance - which required the postulation of a separate {\it strong} 
component in the classical approach (\cite{kap89}) - are naturally produced by AGB stars in the investigated mass range as 
anticipated by Gallino et al. (1998). 
However, a better description of the $s$ abundances is required in the mass region 88 $<$ $A$ $<$ 130 where 
the present yields are somewhat too low. This difference may well be accounted for by the $s$ contributions 
from intermediate mass AGB stars (\cite{vap99}; \cite{gp99}).

\acknowledgements

It is a pleasure to thank F.-K. Thielemann and G. J. Wasserburg for stimulating discussions and suggestions.

This work was partly supported by a grant of italian MURST Cofin 98.

\clearpage

\begin{deluxetable}{ccccc}
\tablenum{1}
\tablewidth{0pc}
\tablecolumns{56}
\tablecaption{MAXWELLIAN AVERAGED NEUTRON CAPTURE CROSS SECTIONS \label{tbl1} }
\tablehead{
\colhead{{\sc Nucleus}} &
\multicolumn{3}{c}{$\left<\sigma\right>$(mbarn)} &
\colhead{{\sc Ref.}} \\
\cline{2-4} \\
%\colhead{} &
%\colhead{} &
%\colhead{} &
%\colhead{} &
\colhead{} &
\colhead{10 keV} &
\colhead{25 keV} &
\colhead{30 keV} &
\colhead{}\\}
\startdata
$^{140}$Ce & \nodata        & 12.0$\pm$0.4   & 11.0$\pm$0.4   & K\"appeler et al. 1996 \nl
           & 16.9$\pm$0.8   & 11.6$\pm$0.6   & 10.6$\pm$0.5   & Beer et al. 1992 \nl
 & & & & \nl
$^{141}$Ce & 186            & 102            & 91.0           & K\"appeler et al. 1996\tablenotemark{a} \nl
           & 357            & \nodata        & 167            & Beer et al. 1992\tablenotemark{a} \nl
 & & & & \nl
$^{142}$Ce & \nodata        & 30.8$\pm$1.0   & 28.3$\pm$1.0   & K\"appeler et al. 1996 \nl
           & \nodata        & \nodata        & 19.6$\pm$1.1   & Beer \& K\"appeler 1980 \nl
 & & & & \nl
$^{141}$Pr & 246.5$\pm$6.5  & 126.3$\pm$1.7  & 111.4$\pm$1.4  & Voss et al. 1999 \nl
           & 196            & 109            & 97             & K\"appeler et al. 1996\tablenotemark{a} \nl
 & & & & \nl
$^{142}$Pr & 684            & 343            & 297            & K\"appeler et al. 1996\tablenotemark{a} \nl
           & \nodata        & \nodata        & 932            & Holmes et al. 1976\tablenotemark{a} \nl
 & & & & \nl
$^{142}$Nd & 65.1$\pm$1.9   & 38.4$\pm$0.8   & 35.0$\pm$0.7   & Wisshak et al. 1998b \nl
           & 65.8$\pm$2.9   & \nodata        & 36.6$\pm$3.0   & Guber et al. 1997 \nl
           & 95.8$\pm$8.3   & 51.9$\pm$4.5   & 46.0$\pm$4.0   & Beer et al. 1992 \nl
 & & & & \nl
$^{143}$Nd & 528.3$\pm$12.0   & 275.7$\pm$3.7  & 244.6$\pm$3.1  & Wisshak et al. 1998a  \nl
           & 508$\pm$21     & 274$\pm$11     & 242$\pm$10     & Beer et al. 1992 \nl
 & & & & \tablebreak
$^{144}$Nd & 147.0$\pm$4.5  & 88.5$\pm$1.7   & 81.3$\pm$1.5   & Wisshak et al. 1998b \nl
           & 122.2$\pm$5.4  & \nodata        & 73.2$\pm$6.1   & Guber et al. 1997 \nl
           & 232$\pm$13     & 123$\pm$7      & 108$\pm$6      & Beer et al. 1992 \nl
\enddata
\tablenotetext{a}{Calculated values}
\end{deluxetable}

\clearpage
\begin{deluxetable}{llccccccccc}                                                                                                                       
\tablenum{2}
\scriptsize                                                                                                                                        
%\tablewidth{0pt}                                                                                                                                      
\tablecolumns{11}                                                                                                                                     
\tablecaption{$S$-PROCESS YIELDS AND RESIDUALS FOR THE STELLAR AND THE CLASSICAL MODEL                                                                
\label{tbl2} }                                                                                                                                        
\tablehead{                                                                                                                                           
\colhead{} &                                                                                                                                          
\colhead{} &                                                                                                                                          
\multicolumn{4}{c}{\sc Stellar model\tablenotemark{a}} &                                                                                                               
\colhead{} &                                                                                                                                          
\multicolumn{4}{c}{\sc Classical model\tablenotemark{a}} \\                                                                                                             
\colhead{{\sc Nucleus}} &
\colhead{{\sc Solar  }} \\                                                                                                                             
\cline{3-6}                                                                                                                                          
\cline{8-11} \\                                                                                                                                        
\colhead{} &                                                                                                                                          
\colhead{{\sc abundance\tablenotemark{b} }} &                                                                                                                                          
\colhead{$N_s^{main}$} &                                                                                                                                     
\colhead{$\delta N_s^{main}$} &                                                                                                                        
\colhead{$N_{r}$} &                                                                                                                                     
\colhead{$\delta N_{r}$} &                                                                                                                        
\colhead{} &                                                                                                                                          
\colhead{$N_s^{main}$} &                                                                                                                                     
\colhead{$\delta N_s^{main}$} &                                                                                                                        
\colhead{$N_{r}$} &                                                                                                                                     
\colhead{$\delta N_{r}$} \\
\colhead{} &                                                                                                                                          
\colhead{} &                                                                                                                                          
\colhead{} &                                                                                                                                          
\colhead{(\%)} &                                                                                                                                          
\colhead{} &                                                                                                                                          
\colhead{(\%)} &                                                                                                                                          
\colhead{} &                                                                                                                                          
\colhead{} &                                                                                                                                          
\colhead{(\%)} &                                                                                                                                          
\colhead{} &                                                                                                                                          
\colhead{(\%)} \\                                                                                                                                          
}
\startdata
$^{ 63}$Cu & 3.61E+02 & 2.95E+00 & 18.6 &	 & &  
&	1.73E+01	&	18.6	&		&	\nl
$^{ 65}$Cu	&	1.61E+02	&	2.04E+00	&	14.5	&	&	& 
&	9.11E+00	&	14.5	&		&	\nl
{\bf Cu}\tablenotemark{d}    &   &    1.0\%                     &           &                 &           & &       5.1\%     
&                  &                  &           \nl 
\tablevspace{10pt} 
$^{ 64}$Zn	&	6.13E+02	&	9.21E-01	&	9.5	&		&	& 
&	4.12E+00	&	9.5	&	&	\nl
$^{ 66}$Zn	&	3.52E+02	&	3.44E+00	&	9.6	&	&		& 
&	1.42E+01	&	9.6	&		&	\nl
$^{ 67}$Zn	&	5.17E+01	&	7.78E-01	&	10.7	&		&	& 
&	3.19E+00	&	10.7	&		&	\nl
$^{ 68}$Zn	&	2.36E+02	&	6.78E+00	&	13.3	&		&	& 
&	2.17E+01	&	13.3	&	 & \nl
$^{ 70}$Zn	&	7.80E+00	&	2.36E-02	&	50.2	&		&	& &
	6.57E-3	&	50.2	&		&	\nl
{\bf Zn}\tablenotemark{d}    &   &       0.9\%                  &           &                 &           & &        3.4\%  &                  
&                  &           \nl
\tablevspace{10pt}
$^{ 69}$Ga	&	2.27E+01	&	8.73E-01	&	11.1	&	&		& 
&	2.85E+00	&	11.1	&	{\it 6.29E+00}\tablenotemark{c}	&	\nl
$^{ 71}$Ga	&	1.51E+01	&	8.35E-01	&	9.0	&		&	& 
&	4.25E+00	&	9.0	&	{\it 0.00E+00}\tablenotemark{c}	&	\nl
{\bf Ga}\tablenotemark{d}    &   &       4.5\%                  &           &                 &           & &       19\%  &                  
&                  &           \nl
\tablevspace{10pt}
$^{ 70}$Ge	&	2.44E+01	&	1.60E+00	&	11.2	&	&	& 
&	4.37E+00	&	11.2	&	&	\nl
$^{ 72}$Ge	&	3.26E+01	&	2.47E+00	&	29.7	&	&	& 
&	6.28E+00	&	29.7	& {\it 8.12E+00}\tablenotemark{c}	&	\nl
$^{ 73}$Ge	&	9.28E+00	&	4.68E-01	&	30.8	&	&	& 
&	1.25E+00	&	30.8	& {\it 3.66E+00}\tablenotemark{c}	&	\nl
$^{ 74}$Ge	&	4.34E+01	&	2.62E+00	&	14.7	&	&	& 
&	6.11E+00	&	14.7	& {\it 2.50E+01}\tablenotemark{c}	&	\nl
$^{ 76}$Ge	&	9.28E+00	&	5.52E-03	&	12.6	&	&	& &	
3.17E-03	&	12.6	&	{\it 9.28E+00}\tablenotemark{c}	&	\nl
{\bf Ge}\tablenotemark{d}    &    &      6.0\%                 &           &                 &           & &       15\%  &                  
&                  &           \nl
\tablevspace{10pt}
$^{ 75}$As	&	6.56E+00	&	3.02E-01	&	12.6	&	&	& &	
5.84E-01	&	12.6	&	{\it 4.44E+00}\tablenotemark{c}	&	\nl
{\bf As}\tablenotemark{d}    &    &      4.6\%                 &           &                 &           & &        8.9\%  &                  
&                  &           \nl
\tablevspace{10pt}
$^{ 76}$Se	&	5.60E+00	&	8.62E-01	&	8.0	&	&	& 
&	2.21E+00	&	8.0	&	&	\nl
$^{ 77}$Se	&	4.70E+00	&	3.15E-01	&	30.2	&	&	& &	
7.27E-01	&	30.2	&	{\it 2.60E+00}\tablenotemark{c}	&	\nl
$^{ 78}$Se	&	1.47E+01	&	1.57E+00	&	21.0	&	&	& 
&	3.44E+00	&	21.0	&	{\it 6.84E+00}\tablenotemark{c}	&	\nl
$^{ 80}$Se	&	3.09E+01	&	2.73E+00	&	9.4	&	&	& 
&	3.49E+00	&	9.4	&	{\it 2.41E+01}\tablenotemark{c}	&	\nl
$^{ 82}$Se	&	5.70E+00	&	3.39E-03	&	50.4	&	&	& &	
1.82E-03	&	50.4	&	{\it 5.70E+00}\tablenotemark{c}	&	\nl
{\bf Se}\tablenotemark{d}    &  &        8.9\%                  &           &                 &           & &       16\%  &                  
&                  &           \nl
\tablevspace{10pt}
$^{ 79}$Br	&	5.98E+00	&	5.22E-01	&	19.4	&	&	& &	
9.58E-01	&	19.4	&	{\it 3.51E+00}\tablenotemark{c}	&	\nl
$^{ 81}$Br	&	5.82E+00	&	5.41E-01	&	19.4	&	&	& &	
5.64E-01	&	19.4	&	{\it 4.36E+00}\tablenotemark{c}	&	\nl
{\bf Br}\tablenotemark{d}    &  &        9.0\%                  &           &                 &           & &        14\%  &                  
&                  &           \tablebreak
\tablevspace{10pt}
$^{ 80}$Kr	&	9.99E-01	&	1.17E-01	&	18.9	&	&	& &	
5.64E-01	&	18.9	&	&	\nl
$^{ 82}$Kr	&	5.15E+00	&	1.91E+00	&	19.6	&	&	& 
&	3.38E+00	&	19.6	&	&	\nl
$^{ 83}$Kr	&	5.16E+00	&	6.50E-01	&	19.1	&	&	& 
&	1.12E+00	&	19.1	&	{\it 3.27E+00}\tablenotemark{c}	&	\nl
$^{ 84}$Kr	&	2.57E+01	&	3.54E+00	&	21.5	&	&	& 
&	8.21E+00	&	21.5	&	{\it 1.68E+01}\tablenotemark{c}	&	\nl
$^{ 86}$Kr	&	7.84E+00	&	2.12E+00	&	18.1	&	&	& &	{\bf 
64\%}\tablenotemark{e}	&	18.1	&	        	&	   	\nl
{\bf Kr}\tablenotemark{d}    &   &      19\%                   &           &                 &           & &       47\%  &                  
&                  &           \nl
\tablevspace{10pt}
$^{ 85}$Rb	&	5.12E+00	&	8.36E-01	&	7.6	&	&	& 
&	2.14E+00	&	7.6	&	{\it 2.75E+00}\tablenotemark{c}	&	\nl
$^{ 87}$Rb	&	2.11E+00	&	7.46E-01	&	11.7	&	&	& &	{\bf 
66\%}\tablenotemark{e}	&	11.7	&	        	&	   	\nl
{\bf Rb}\tablenotemark{d}   & &         22\%                    &           &                 &           & &       59\%  &                  
&                  &           \nl
\tablevspace{10pt}
$^{ 86}$Sr	&	2.32E+00	&	1.09E+00	&	8.8	&	&	& 
&	1.58E+00	&	8.8	&		&	\nl
$^{ 87}$Sr	&	1.51E+00	&	7.60E-01	&	8.9	&	&	& 
&	1.12E+00	&	8.9	&	&	\nl
$^{ 88}$Sr	&	1.94E+01	&	1.79E+01	&	8.1	&	&	& 
&	1.82E+01	&	8.1	&	{\it 9.06E-01}\tablenotemark{c}	&	\nl
{\bf Sr}\tablenotemark{d}    &    &     85\%                    &           &                 &           & &       90\%  &                  
&                  &           \nl
\tablevspace{10pt}
$^{ 89}$Y	&	4.64E+00	&	4.27E+00	&	6.6	&	3.72E-01	&	100	& &	{\bf 
6.4\%}\tablenotemark{e}	&	6.6	&	        	&	   	\nl
{\bf Y}\tablenotemark{d}     &   & 92\%  &    &   &  & &   100\%  & & & \nl
\tablevspace{10pt}
$^{ 90}$Zr	&	5.87E+00	&	4.24E+00	&	12.4	&	1.63E+00	&	39.7	& 
&	4.01E+00	&	12.4	&	1.86E+00	&	33.5	\nl
$^{ 91}$Zr	&	1.28E+00	&	1.23E+00	&	14.8	&	5.46E-02	&	100	& &	{\bf 
0.50\%}\tablenotemark{e}	&	14.8	&	        	&	   	\nl
$^{ 92}$Zr	&	1.96E+00	&	1.83E+00	&	13.7	&	1.28E-01	&	100	& &	{\bf 
8.3\%}\tablenotemark{e}	&	13.7	&	        	&	   	\nl
$^{ 94}$Zr	&	1.98E+00	&	{\bf 8.2\%}\tablenotemark{e}	&	7.3	&	        	&	   	& &	{\bf 
16\%}\tablenotemark{e}	&	7.3	&	        	&	   	\nl
$^{ 96}$Zr	&	3.20E-01	&	1.76E-01	&	7.3	&		&	& &	
1.63E-01	&	7.3	&	&	\nl
{\bf Zr}\tablenotemark{d}    &    &     83\%                    &           &                 &           & &       82\%  &                  
&                  &           \nl
\tablevspace{10pt}
$^{ 93}$Nb	&	6.98E-01	&	5.96E-01	&	2.4	&	1.02E-01	&	16.8	& &	{\bf 2.0\%}\tablenotemark{e}
&	2.4	&		&		\nl
{\bf Nb}\tablenotemark{d}    &      &   85\%                    &           &                 &           & &        100\%  &                  
&                  &           \nl
\tablevspace{10pt}
$^{ 94}$Mo	&	2.36E-01	&	1.53E-03	&	20.0	&	&	& &	
8.27E-05	&	20.0	&	&	\nl
$^{ 95}$Mo	&	4.06E-01	&	2.25E-01	&	6.9	&	1.81E-01	&	15.0	& &	
2.24E-01	&	6.9	&	1.82E-01	&	14.9	\nl
$^{ 96}$Mo	&	4.25E-01	&	{\bf 6.1\%}\tablenotemark{e}	&	7.1	&	        	&	   	& &	{\bf 
16\%}\tablenotemark{e}	&	7.1	&	        	&	   	\nl
$^{ 97}$Mo	&	2.44E-01	&	1.43E-01	&	6.9	&	1.01E-01	&	16.4	& &	
1.67E-01	&	6.9	&	7.72E-02	&	22.9	\nl
$^{ 98}$Mo	&	6.15E-01	&	4.66E-01	&	7.5	&	1.49E-01	&	32.7	& &	
5.52E-01	&	7.5	&	6.25E-02	&	85.3	\nl
$^{100}$Mo	&	2.46E-01   &   9.42E-03  &     21.3  &    2.37E-01  &      5.8  & &    0.00E-00  &     21.3 &     2.46E-01     
&   5.5 \nl
{\bf Mo}\tablenotemark{d}    &     &      50\%                  &           &                 &           & &       54\%  &                  
&                  &           \tablebreak
\tablevspace{10pt}
$^{ 99}$Ru	&	2.36E-01	&	6.69E-02	&	50.3	&	1.69E-01	&	21.3	& &	
6.94E-02	&	50.3	&	1.67E-01	&	22.3	\nl
$^{100}$Ru	&	2.34E-01	&	2.23E-01	&	8.3	&	&	& &	{\bf 
11\%}\tablenotemark{e}	&	8.3	&	        	&	   	\nl
$^{101}$Ru	&	3.16E-01	&	4.83E-02	&	6.7	&	2.68E-01	&	6.5	& &	
5.38E-02	&	6.7	&	2.62E-01	&	6.7	\nl
$^{102}$Ru	&	5.88E-01	&	2.53E-01	&	8.0	&	3.35E-01	&	11.3	& &	
2.82E-01	&	8.0	&	3.06E-01	&	12.8	\nl
$^{104}$Ru	&	3.48E-01	&	9.52E-03	&	8.2	&	3.38E-01	&	5.6	& &	
3.75E-03	&	8.2	&	3.44E-01	&	5.5	\nl
{\bf Ru}\tablenotemark{d}    &   &   32\%                       &           &                 &           & &       37\%  &                  
&                  
&           \nl
\tablevspace{10pt}
$^{103}$Rh	&	3.44E-01	&	4.67E-02	&	8.2	&	2.97E-01	&	9.3	& &	
6.50E-02	&	8.2	&	2.79E-01	&	10.0	\nl
{\bf Rh}\tablenotemark{d}    &   &    14\%                      &           &                 &           & &       19\%  &                  
&                  &           \nl
\tablevspace{10pt}
$^{104}$Pd	&	1.55E-01	&	{\bf 5.7\%}\tablenotemark{e}	&	12.0	&	        	&	   	& &	{\bf 
14\%}\tablenotemark{e}	&	12.0	&	        	&	   	\nl
$^{105}$Pd	&	3.10E-01	&	4.27E-02	&	8.3	&	2.67E-01	&	7.8	& &	
4.31E-02	&	8.3	&	2.67E-01	&	7.8	\nl
$^{106}$Pd	&	3.80E-01	&	1.95E-01	&	11.9	&	1.85E-01	&	18.5	& &	
2.02E-01	&	11.9	&	1.78E-01	&	19.5	\nl
$^{108}$Pd	&	3.68E-01	&	2.40E-01	&	11.9	&	1.28E-01	&	29.2	& &	
2.46E-01	&	11.9	&	1.22E-01	&	31.1	\nl
$^{110}$Pd	&	1.63E-01	&	5.93E-03	&	15.2	&	1.57E-01	&	6.9	& &	
2.64E-04	&	15.2	&	1.63E-01	&	6.6	\nl
{\bf Pd}\tablenotemark{d}    & & 46\%                          &           &                 &           & &       47\%  &                  
&                  &           \nl
\tablevspace{10pt}
$^{107}$Ag	&	2.52E-01	&	3.77E-02	&	4.2	&	2.14E-01	&	3.5	& &	
3.80E-02	&	4.2	&	2.14E-01	&	3.5	\nl
$^{109}$Ag	&	2.34E-01	&	5.86E-02	&	4.1	&	1.75E-01	&	4.1	& &	
6.44E-02	&	4.1	&	1.70E-01	&	4.3	\nl
{\bf Ag}\tablenotemark{d}    & &  20\%                        &           &                 &           & &       21\%  &                  
&                  &           \nl
\tablevspace{10pt}
$^{108}$Cd	&	1.43E-02	&	1.61E-05	&	18.4	&	&	& &	
2.20E-04	&	18.4	&	&	\nl
$^{110}$Cd	&	2.01E-01	&	1.95E-01	&	13.8	&	&	& &	
2.00E-01	&	13.8	&	&	\nl
$^{111}$Cd	&	2.06E-01	&	4.87E-02	&	13.4	&	1.57E-01	&	9.5	& &	
4.61E-02	&	13.4	&	1.60E-01	&	9.2	\nl
$^{112}$Cd	&	3.88E-01	&	2.05E-01	&	14.4	&	1.83E-01	&	21.2	& &	
2.05E-01	&	14.4	&	1.83E-01	&	21.3	\nl
$^{113}$Cd	&	1.97E-01	&	6.85E-02	&	12.8	&	1.29E-01	&	12.1	& &	
7.23E-02	&	12.8	&	1.25E-01	&	12.7	\nl
$^{114}$Cd	&	4.63E-01	&	2.95E-01	&	16.8	&	1.68E-01	&	34.5	& &	
3.68E-01	&	16.8	&	9.45E-02	&	73.0	\nl
$^{116}$Cd	&	1.21E-01	&	2.13E-02	&	14.3	&	9.97E-02	&	8.5	& &	
9.16E-03	&	14.3	&	1.12E-01	&	7.1	\nl
{\bf Cd}\tablenotemark{d}    &  &  52\%                         &           &                 &           & &       57\%  &                  
&                  &           \nl
\tablevspace{10pt}
$^{113}$In	&	7.90E-03	&	5.59E-08	&	11.0	&	&	& &	
3.63E-05	&	11.0	&	&	\nl
$^{115}$In	&	1.76E-01	&	6.43E-02	&	11.8	&	1.12E-01	&	12.2	& &	
6.66E-02	&	11.8	&	1.09E-01	&	12.6	\nl
{\bf In}\tablenotemark{d}    &  & 35\%                          &           &                 &           & &       36\%  &                  
&                  &           \tablebreak
\tablevspace{10pt}
$^{114}$Sn	&	2.52E-02	&	4.75E-06	&	9.5	&	&	& &	
2.00E-04	&	9.5	&	&	\nl
$^{115}$Sn	&	1.29E-02	&	3.06E-04	&	9.7	&	&	& &	
1.60E-04	&	9.7	&	&	\nl
$^{116}$Sn	&	5.55E-01	&	4.76E-01	&	9.5	&	&	& &	
4.89E-1	&	9.5	&	&	\nl
$^{117}$Sn	&	2.93E-01	&	1.41E-01	&	9.5	&	1.52E-01	&	20.2	& &	
1.40E-01	&	9.5	&	1.53E-01	&	20.0	\nl
$^{118}$Sn	&	9.25E-01	&	6.67E-01	&	9.4	&	2.58E-01	&	41.7	& &
6.86E-01	&	9.4	&	2.39E-01	&	45.5	\nl
$^{119}$Sn	&	3.28E-01	&	1.27E-01	&	21.2	&	2.01E-01	&	20.5	& &	
2.31E-01	&	21.2	&	9.73E-02	&	59.5	\nl
$^{120}$Sn	&	1.25E+00	&	9.77E-01	&	9.5	&	2.68E-01	&	55.7	& 
&	1.07E+00	&	9.5	&	1.78E-01	&	86.7	\nl
$^{122}$Sn	&	1.77E-01	&	7.93E-02	&	26.7	&	9.77E-02	&	27.6	& &	
1.54E-02	&	26.7	&	1.62E-01	&	10.6	\nl
{\bf Sn}\tablenotemark{d}    &   &  65\%                        &           &                 &           & &       70\%  &                  
&                  &           \nl
\tablevspace{10pt}
$^{121}$Sb	&	1.77E-01	&	6.78E-02	&	18.2	&	1.09E-01	&	31.3	& &	
7.22E-02	&	18.2	&	1.05E-01	&	32.9	\nl
$^{123}$Sb	&	1.32E-01	&	8.06E-03	&	18.2	&	1.24E-01	&	19.2	& &	
2.19E-03	&	18.2	&	1.30E-01	&	18.3	\nl
{\bf Sb}\tablenotemark{d}    &    &  25\%                       &           &                 &           & &       24\%  &                  
&                  &           \nl
\tablevspace{10pt}
$^{122}$Te	&	1.24E-01	&	1.09E-01	&	10.0	&	&	& & {\bf 
0.51\%}\tablenotemark{e}	&	10.0	&	        	&	   	\nl
$^{123}$Te	&	4.28E-02	&	3.83E-02	&	10.0	&	&	& &	{\bf 
2.9\%}\tablenotemark{e}	&	10.0	&	        	&	   	\nl
$^{124}$Te	&	2.29E-01	&	2.08E-01	&	10.1	&	&	& &	{\bf 
3.1\%}\tablenotemark{e}	&	10.1	&	        	&	   	\nl
$^{125}$Te	&	3.42E-01	&	6.80E-02	&	10.0	&	2.74E-01	&	12.7	& &	
8.41E-02	&	10.0	&	2.58E-01	&	13.7	\nl
$^{126}$Te	&	9.09E-01	&	3.68E-01	&	10.1	&	5.41E-01	&	18.2	& &	
4.29E-01	&	10.1	&	4.80E-01	&	21.0	\nl
$^{128}$Te	&	1.53E+00	&	2.47E-02	&	10.4	&	1.50E+00	&	10.2	& &	
4.03E-03	&	10.4	&	1.52E+00	&	10.0	\nl
{\bf Te}\tablenotemark{d}    &   &   17\%                       &           &                 &           & &       19\%  &                  
&                  &           \nl
\tablevspace{10pt}
$^{127}$I 	&	9.00E-01	&	4.75E-02	&	21.5	&	8.53E-01	&	22.2	& &	
5.50E-02	&	21.5	&	8.45E-01	&	22.4	\nl
{\bf I}\tablenotemark{d}     &  &   5.3\%                       &           &                 &           & &        6.1\%  &                  
&                  &           \nl
\tablevspace{10pt}
$^{128}$Xe	&	1.03E-01	&	8.42E-02	&	37.7	&	&	& &	
9.85E-02	&	37.7	&	&	\nl
$^{129}$Xe	&	1.28E+00	&	4.03E-02	&	21.8	&	1.24E+00	&	20.7	& &	
4.57E-02	&	21.8	&	1.23E+00	&	20.8	\nl
$^{130}$Xe	&	2.05E-01	&	1.70E-01	&	34.4	&	&	& &	
1.92E-01	&	34.4	&		&	\nl
$^{131}$Xe	&	1.02E+00	&	6.55E-02	&	26.8	&	9.55E-01	&	21.5	& &	
7.39E-02	&	26.8	&	9.46E-01	&	21.7	\nl
$^{132}$Xe	&	1.24E+00	&	4.16E-01	&	21.6	&	8.24E-01	&	32.0	& &	
4.92E-01	&	21.6	&	7.48E-01	&	36.1	\nl
$^{134}$Xe	&	4.59E-01	&	2.22E-02	&	21.7	&	4.37E-01	&	21.0	& &	
8.26E-03	&	21.7	&	4.51E-01	&	20.4	\nl
{\bf Xe}\tablenotemark{d}    &  &  17\%                        &           &                 &           & &       20\%  &                  
&                  &           \nl
\tablevspace{10pt}
$^{133}$Cs	&	3.72E-01	&	5.39E-02	&	7.0	&	3.18E-01	&	6.7	& &	
6.31E-02	&	7.0	&	3.09E-01	&	6.9	\nl
{\bf Cs}\tablenotemark{d}    &  &  15\%                        &           &                 &           & &       17\%  &                  
&                  &           \tablebreak
\tablevspace{10pt}
$^{134}$Ba	&	1.09E-01	&	1.07E-01	&	7.1	&	&	& &	{\bf 
58\%}\tablenotemark{e}	&	7.1	&	        	&	   	\nl
$^{135}$Ba	&	2.96E-01	&	7.75E-02	&	7.1	&	2.19E-01	&	8.9	& &	
8.05E-02	&	7.1	&	2.15E-01	&	9.1	\nl
$^{136}$Ba	&	3.53E-01	& {\bf 0.25\%}\tablenotemark{e}	&	7.2	&	        	&	   	& &	{\bf 
37\%}\tablenotemark{e}	&	7.2	&	        	&	   	\nl
$^{137}$Ba	&	5.04E-01	&	3.30E-01	&	7.3	&	1.74E-01	&	23.0	& &	
3.67E-01	&	7.3	&	1.37E-01	&	30.4	\nl
$^{138}$Ba	&	3.22E+00	&	2.76E+00	&	6.5	&	4.59E-01	&	58.8	& &	{\bf 
18\%}\tablenotemark{e}	&	6.5	&	        	&	   	\nl
{\bf Ba}\tablenotemark{d}    &  & 81\%                          &           &                 &           & &       92\%  &                  
&                  &           \nl
\tablevspace{10pt}
$^{139}$La	&	4.46E-01	&	2.77E-01	&	7.3	&	1.69E-01	&	13.1	& &
	3.69E-01	&	7.3	&	7.71E-02	&	36.8	\nl
{\bf La}\tablenotemark{d}    &  & 62\%                         &           &                 &           & &       83\%  &                  
&                  &           \nl
\tablevspace{10pt}
$^{140}$Ce	&	1.00E+00	&	8.36E-01	&	4.0	&	1.69E-01	&	22.3	& &	
9.68E-01	&	4.0	&	3.66E-02	&	100	\nl
$^{142}$Ce	&	1.26E-01	&	2.79E-02	&	3.9	&	9.81E-02	&	2.5	& &	
1.17E-02	&	3.9	&	1.14E-01	&	1.9	\nl
{\bf Ce}\tablenotemark{d}    & &  77\%                          &           &                 &           & &       87\%  &                  
&                  &           \nl
\tablevspace{10pt}
$^{141}$Pr	&	1.67E-01	&	8.13E-02	&	2.7	&	8.57E-02	&	5.3	& &	
9.33E-02	&	2.7	&	7.37E-02	&	6.4	\nl
{\bf Pr}\tablenotemark{d}    &  & 49\%                         &           &                 &           & &       56\%  &                  
&                  &           \nl
\tablevspace{10pt}
$^{142}$Nd	&	2.25E-01	&	2.08E-01	&   2.4	&	&	& &	{\bf 
12\%}\tablenotemark{e}	&	2.4	&	        	&	   	\nl
$^{143}$Nd	&	1.00E-01	&	3.16E-02	&	1.9	&	6.84E-02	&	2.1	& &	
3.75E-02	&	1.9	&	6.25E-02	&	2.4	\nl
$^{144}$Nd	&	1.97E-01	&	1.00E-01	&	2.2	&	9.68E-02	&	3.5	& &	
1.08E-01	&	2.2	&	8.93E-02	&	3.9	\nl
$^{145}$Nd	&	6.87E-02	&	1.89E-02	&	1.7	&	4.98E-02	&	1.9	& &	
2.20E-02	&	1.7	&	4.67E-02	&	2.1	\nl
$^{146}$Nd	&	1.42E-01	&	9.11E-02	&	1.7	&	5.09E-02	&	4.7	& &	
9.12E-02	&	1.7	&	5.08E-02	&	4.7	\nl
$^{148}$Nd	&	4.77E-02	&	9.05E-03	&	1.8	&	3.87E-02	&	1.7	& &	
3.77E-03	&	1.8	&	4.39E-02	&	1.4	\nl
{\bf Nd}\tablenotemark{d}    &  & 56\%                         &           &                 &           & &       59\%  &                  
&                  &           \nl
%\tablevspace{10pt}
$^{147}$Sm	&	3.99E-02	&	8.25E-03	&	1.7	&	3.17E-02	&	1.7	& &
8.91E-03	&	1.7	&	3.10E-02	&	1.7	\nl
$^{148}$Sm	&	2.92E-02	&	2.82E-02	&	1.6	&	&	& &	{\bf 
1.8\%}\tablenotemark{e}	&	1.6	&	        	&	   	\nl
$^{149}$Sm	&	3.56E-02	&	4.45E-03	&	1.6	&	3.12E-02	&	1.5	& &	
4.49E-03	&	1.6	&	3.11E-02	&	1.5	\nl
$^{150}$Sm	&	1.91E-02	&	1.91E-02	&	1.6	&	        	&	   	& & 	
1.91E-02	&	1.6	&	        	&	   	\nl
$^{152}$Sm	&	6.89E-02	&	1.58E-02	&	1.6	&	5.31E-02	&	1.8	& &	
1.55E-02	&	1.6	&	5.34E-02	&	1.7	\nl
$^{154}$Sm	&	5.86E-02	&	4.69E-04	&	6.6	&	5.81E-02	&	1.3	& &	
3.18E-04	&	6.6	&	5.83E-02	&	1.3	\nl
{\bf Sm}\tablenotemark{d}    &  & 29\%                          &           &                 &           & &       30\%  &                  
&                  &           \nl
\tablevspace{10pt}
$^{151}$Eu	&	4.65E-02	&	3.04E-03	&	4.3	&	4.35E-02	&	1.7	& &	
4.29E-03	&	4.3	&	4.22E-02	&	1.8	\nl
$^{153}$Eu	&	5.08E-02	&	2.58E-03	&	10.1	&	4.82E-02	&	1.8	& &	
3.04E-03	&	10.1	&	4.78E-02	&	1.8	\nl
{\bf Eu}\tablenotemark{d}    &  &  5.8\%                       &           &                 &           & &        7.5\%  &                  
&                  &           \tablebreak
\tablevspace{10pt}
$^{152}$Gd\tablenotemark{g}	&	6.60E-04	&	5.83E-04	&	2.1	&	&	& &	
4.72E-04	&	2.1	&	&	\nl
$^{154}$Gd	&	7.19E-03	&	6.85E-03	&	1.8	&	&	& &	
6.26E-03	&	1.8	&	&	\nl
$^{155}$Gd	&	4.88E-02	&	2.88E-03	&	1.8	&	4.59E-02	&	1.5	& &	
3.50E-03	&	1.8	&	4.53E-02	&	1.5	\nl
$^{156}$Gd	&	6.76E-02	&	1.15E-02	&	1.6	&	5.61E-02	&	1.7	& &	
1.28E-02	&	1.6	&	5.48E-02	&	1.8	\nl
$^{157}$Gd	&	5.16E-02	&	5.53E-03	&	1.8	&	4.61E-02	&	1.6	& &	
5.76E-03	&	1.8	&	4.58E-02	&	1.6	\nl
$^{158}$Gd	&	8.20E-02	&	2.25E-02	&	1.6	&	5.95E-02	&	2.0	& &	
2.40E-02	&	1.6	&	5.80E-02	&	2.1	\nl
$^{160}$Gd	&	7.21E-02	&	8.27E-04	&	9.7	&	7.13E-02	&	1.4	& &
	2.22E-05	&	9.7	&	7.21E-02	&	1.4	\nl
{\bf Gd}\tablenotemark{d}    &  & 15\%                          &           &                 &           & &       16\%  &                  
&                  &           \nl
\tablevspace{10pt}
$^{159}$Tb	&	6.03E-02	&	4.36E-03	&	6.0	&	5.59E-02	&	2.4	& &	
5.05E-03	&	6.0	&	5.52E-02	&	2.5	\nl
{\bf Tb}\tablenotemark{d}    &  &  7.2\%                  &           &                 &                & &        8.4\%  &                  
&                  &           \nl
\tablevspace{10pt}
$^{160}$Dy	&	9.22E-03	&	8.06E-03	&	1.9	&	&	& &	
8.34E-03	&	1.9	&	&	\nl
$^{161}$Dy	&	7.45E-02	&	4.12E-03	&	1.7	&	7.04E-02	&	1.5	& &	
3.88E-03	&	1.7	&	7.06E-02	&	1.5	\nl
$^{162}$Dy	&	1.01E-01	&	1.64E-02	&	1.6	&	8.46E-02	&	1.7	& &	
1.66E-02	&	1.6	&	8.44E-02	&	1.7	\nl
$^{163}$Dy	&	9.82E-02	&	3.52E-03	&	2.0	&	9.47E-02	&	1.5	& &
	4.85E-03	&     2.0	&	9.34E-02	&	1.5	\nl
$^{164}$Dy	&	1.11E-01	&	2.61E-02	&	1.9	&	8.49E-02	&	1.9	& &	
1.57E-02	&	1.9	&	9.53E-02	&	1.7	\nl
{\bf Dy}\tablenotemark{d}    & &  15\%                           &           &                 &           & &       13\%  &                  
&                  &           \nl
\tablevspace{10pt}
$^{165}$Ho	&	8.89E-02	&	6.95E-03	&	5.6	&	8.20E-02	&	2.6	& &	
5.82E-03	&	5.6	&	8.31E-02	&	2.6	\nl
{\bf Ho}\tablenotemark{d}    &  &  7.8\%                        &           &                 &           & &        6.5\%  &                  
&                  &           \nl
\tablevspace{10pt}
$^{164}$Er	&	4.04E-03	&	3.34E-03	&	4.9	&	&	& &
	3.49E-03	&	4.9	&	&	\nl
$^{166}$Er	&	8.43E-02	&	1.25E-02	&	10.1	&	7.18E-02	&	2.3	& &
	1.34E-02	&	10.1	&	7.09E-02	&	2.5	\nl
$^{167}$Er	&	5.76E-02	&	4.92E-03	&	10.1	&	5.27E-02	&	1.7	& &
	5.23E-03	&	10.1	&	5.24E-02	&	1.7	\nl
$^{168}$Er	&	6.72E-02	&	1.90E-02	&	12.1	&	4.82E-02	&	5.1	& &
	2.17E-02	&	12.1	&	4.55E-02	&	6.1	\nl
$^{170}$Er	&	3.74E-02	&	2.69E-03	&	14.9	&	3.47E-02	&	1.8	& &
	2.19E-03	&	14.9	&	3.52E-02	&	1.7	\nl
{\bf Er}\tablenotemark{d}    & &  17\%                          &           &                 &           & &       18\%  &                  
&                  &           \nl
\tablevspace{10pt}
$^{169}$Tm	&	3.78E-02	&	5.03E-03	&	5.5	&	3.28E-02	&	2.8	& &
	6.78E-03	&	5.5	&	3.10E-02	&	3.0	\nl
{\bf Tm}\tablenotemark{d}    &  & 13\%                          &           &                 &           & &       18\%  &                  
&                  &           \tablebreak
\tablevspace{10pt}
$^{170}$Yb	&	7.56E-03	&	{\bf 1.1\%}\tablenotemark{e}	&	4.2	&	        	&	   	& &
	6.56E-03	&	4.2	&	&	\nl
$^{171}$Yb	&	3.54E-02	&	4.93E-03	&	3.9	&	3.05E-02	&	2.0	& &
	6.07E-03	&	3.9	&	2.93E-02	&	2.1	\nl
$^{172}$Yb	&	5.43E-02	&	1.65E-02	&	8.4	&	3.78E-02	&	4.3	& &	
1.77E-02	&	8.4	&	3.66E-02	&	4.7	\nl
$^{173}$Yb	&	4.00E-02	&	8.57E-03	&	8.5	&	3.14E-02	&	3.1	& &
	8.92E-03	&	8.5	&	3.11E-02	&	3.2	\nl
$^{174}$Yb	&	7.88E-02	&	3.91E-02	&	9.2	&	3.97E-02	&	9.6	& &	
4.27E-02	&	9.2	&	3.61E-02	&	11.5	\nl
$^{176}$Yb	&	3.15E-02	&	4.28E-03	&	10.0	&	2.72E-02	&	2.4	& &	
1.10E-03	&	10.0	&	3.04E-02	&	1.7	\nl
{\bf Yb}\tablenotemark{d}    &  & 33\%                          &           &                 &           & &       34\%  &                  
&                  &           \nl
\tablevspace{10pt}
$^{175}$Lu	&	3.57E-02	&	6.33E-03	&	4.0	&	2.94E-02	&	1.8	& &
	5.91E-03	&	4.0	&	2.98E-02	&	1.7	\nl
$^{176}$Lu	&	1.03E-03	&	{\bf 25\%}\tablenotemark{e}	&	4.1	&	        	&	   	& &	{\bf 
83\%}\tablenotemark{e}	&	4.1	&	        	&	   	\nl
{\bf Lu}\tablenotemark{d}    &  & 20\%                          &           &                 &           & &       19\%  &                  
&                  &           \nl
%\tablevspace{10pt}
$^{176}$Hf	&	7.93E-03	&	7.65E-03	&	4.8	&	&	& &	{\bf
6.9\%}\tablenotemark{e}	&	4.8	&	        	&	   	\nl
$^{177}$Hf	&	2.87E-02	&	5.29E-03	&	4.9	&	2.34E-02	&	2.6	& &	
5.14E-03	&	4.9	&	2.36E-02	&	2.5	\nl
$^{178}$Hf	&	4.20E-02	&	2.40E-02	&	3.7	&	1.80E-02	&	6.7	& &	
2.19E-02	&	3.7	&	2.01E-02	&	5.7	\nl
$^{179}$Hf	&	2.10E-02	&	7.74E-03	&	3.6	&	1.33E-02	&	3.7	& &	
6.80E-03	&	3.6	&	1.42E-02	&	3.3	\nl
$^{180}$Hf	&	5.41E-02	&	4.08E-02	&	3.4	&	1.33E-02	&	13.1	& &	
3.71E-02	&	3.4	&	1.70E-02	&	9.6	\nl
{\bf Hf}\tablenotemark{d}    & & 56\%                          &           &                 &           & &       51\%  &                  
&                  &           \nl
\tablevspace{10pt}
$^{180}$Ta	&	2.48E-06	&	1.21E-06	&	20.1	&	&	& &	{\bf 
17\%}\tablenotemark{e}	&	20.1	&	        	&	   	\nl
$^{181}$Ta	&	2.07E-02	&	8.55E-03	&	2.7	&	1.22E-02	&	3.6	& &	
8.53E-03	&	2.7	&	1.22E-02	&	3.6	\nl
{\bf Ta}\tablenotemark{d}    &  &  41\%                       &           &                 &           & &       41\%  &                  
&                  &           \nl
\tablevspace{10pt}
$^{180}$W 	&	1.73E-04	&	8.02E-06	&	11.6	&		&	& &	
1.38E-04	&	11.6	&	&	\nl
$^{182}$W 	&	3.50E-02	&	1.60E-02	&	7.2	&	1.90E-02	&	11.2	& &	
1.35E-02	&	7.2	&	2.15E-02	&	9.5	\nl
$^{183}$W	&	1.90E-02	&	1.02E-02	&	7.2	&	8.76E-03	&	13.9	& &	
1.24E-02	&	7.2	&	6.64E-03	&	19.8	\nl
$^{184}$W	&	4.08E-02	&	2.88E-02	&	7.3	&	1.20E-02	&	24.5	& &	
2.82E-02	&	7.3	&	1.26E-02	&	23.2	\nl
$^{186}$W   &	3.80E-02	&	1.91E-02	&	7.0	&	1.89E-02	&	12.5	& &
	8.41E-03	&	7.0	&	2.96E-02	&	6.8	\nl
{\bf W}\tablenotemark{d}     &  &  56\%                         &           &                 &           & &       47\%  &                  
&                  &           \nl
\tablevspace{10pt}
$^{185}$Re	&	1.93E-02	&	4.78E-03	&	10.2	&	1.45E-02	&	12.9	& &	
5.91E-03	&	10.2	&	1.34E-02	&	14.3	\nl
$^{187}$Re	&	3.51E-02	&	6.65E-05	&	10.6	&	3.50E-02	&	9.4	& &	
1.55E-03	&	10.6	&	3.36E-02	&	9.8	\nl
{\bf Re}\tablenotemark{d}    &  &  8.9\%                        &           &                 &           & &       14\%  &                  
&                  &           \tablebreak
\tablevspace{10pt}
$^{186}$Os	&	1.07E-02	&	1.04E-02	&	7.4	&	&	& &
	1.06E-02	&	7.4	&	&	\nl
$^{187}$Os	&	8.07E-03	&	6.58E-03	&	7.1	&	&	& &
	4.99E-03	&	7.1	&	&	\nl
$^{188}$Os	&	8.98E-02	&	1.72E-02	&	7.3	&	7.26E-02	&	8.0	& &
	1.54E-02	&	7.3	&	7.44E-02	&	7.7	\nl
$^{189}$Os	&	1.09E-01	&	4.70E-03	&	7.5	&	1.04E-01	&	6.6	& &
	5.25E-03	&	7.5	&	1.04E-01	&	6.6	\nl
$^{190}$Os	&	1.78E-01	&	2.14E-02	&	16.5	&	1.57E-01	&	7.5	& &
	2.04E-02	&	16.5	&	1.58E-01	&	7.4	\nl
$^{192}$Os	&	2.77E-01	&	2.86E-03	&	15.8	&	2.74E-01	&	6.4	& &
	1.03E-03	&	15.8	&	2.76E-01	&	6.3	\nl
{\bf Os}\tablenotemark{d}    &  &  9.4\%                        &           &                 &           & &        8.6\%  &                  
&                  &           \nl
\tablevspace{10pt}
$^{191}$Ir	&	2.47E-01	&	4.68E-03	&	7.9	&	2.42E-01	&	6.2	& &
	4.57E-03	&	7.9	&	2.42E-01	&	6.2	\nl
$^{193}$Ir	&	4.14E-01	&	4.40E-03	&	7.9	&	4.10E-01	&	6.2	& &	
6.13E-03	&	7.9	&	4.08E-01	&	6.2	\nl
{\bf Ir}\tablenotemark{d}    &  &  1.4\%                     &           &                 &           & &        1.6\%  &                  
&                  &           \nl
\tablevspace{10pt}
$^{192}$Pt	&	1.05E-02	&	1.03E-02	&	50.5	&	&	& &	{\bf 
51\%}\tablenotemark{e}	&	50.5	&	        	&	   	\nl
$^{194}$Pt	&	4.41E-01	&	1.77E-02	&	50.5	&	4.23E-01	&	8.0	& &	
1.87E-02	&	50.5	&	4.22E-01	&	8.0	\nl
$^{195}$Pt	&	4.53E-01	&	7.53E-03	&	50.5	&	4.45E-01	&	7.6	& &
	5.58E-03	&	50.5	&	4.47E-01	&	7.5	\nl
$^{196}$Pt	&	3.38E-01	&	3.30E-02	&	13.8	&	3.05E-01	&	8.3	& &
	2.95E-02	&	13.8	&	3.09E-01	&	8.2	\nl
$^{198}$Pt	&	9.63E-02	&	2.33E-05	&	12.1	&	9.63E-02	&	7.4	& &	
5.98E-05	&	12.1	&	9.62E-02	&	7.4	\nl
{\bf Pt}\tablenotemark{d}    &  &  5.1\%                        &           &                 &           & &        4.8\%  &                  
&                  &           \nl
\tablevspace{10pt}
$^{197}$Au	&	1.87E-01	&	1.09E-02	&	15.1	&	1.76E-01	&	16.0	& &	
9.72E-03	&	15.1	&	1.77E-01	&	15.8	\nl
{\bf Au}\tablenotemark{d}    &  &  5.8\%                        &           &                 &           & &        5.2\%  &                  
&                  &           \nl
\tablevspace{10pt}
$^{198}$Hg	&	3.39E-02	&	{\bf 2.4\%}\tablenotemark{e}	&	14.8	&	        	&	   	& &	
3.12E-02	&	14.8	&	&	\nl
$^{199}$Hg	&	5.74E-02	&	1.52E-02	&	13.5	&	4.22E-02	&	17.0	& &	
1.44E-02	&	13.5	&	4.30E-02	&	16.6	\nl
$^{200}$Hg	&	7.85E-02	&	5.15E-02	&	15.9	&	2.70E-02	&	46.3	& &	
4.50E-02	&	15.9	&	3.35E-02	&	35.3	\nl
$^{201}$Hg	&	4.48E-02	&	2.22E-02	&	13.1	&	2.26E-02	&	27.1	& &
	1.93E-02	&	13.1	&	2.55E-02	&	23.3	\nl
$^{202}$Hg	&	1.02E-01	&	8.23E-02	&	14.5	&	1.92E-02	&	88.6	& &
	6.44E-02	&	14.5	&	3.71E-02	&	41.4	\nl
$^{204}$Hg	&	2.33E-02	&	2.07E-03	&	15.3	&	2.12E-02	&	13.3	& &
	3.51E-04	&	15.3	&	2.29E-02	&	12.2	\nl
{\bf Hg}\tablenotemark{d}    & &  61\%                         &           &                 &           & &       51\%  &                  
&                  &           \nl
\tablevspace{10pt}
$^{203}$Tl	&	5.43E-02	&	4.06E-02	&	11.4	&	1.37E-02	&	50.3	& &	
3.70E-02	&	11.4	&	1.73E-02	&	38.3	\nl
$^{205}$Tl	&	1.30E-01	&	9.89E-02	&	12.0	&	3.08E-02	&	55.1	& &	
7.61E-02	&	12.0	&	5.36E-02	&	28.4	\nl
{\bf Tl}\tablenotemark{d}    & &  76\%                         &           &                 &           & &       61\%  &                  
&                  &           \tablebreak
%\tablevspace{10pt}
$^{204}$Pb	&	6.11E-02	&	5.76E-02	&	9.9	&	&	& &
4.83E-02	&	9.9	&	&	\nl
$^{206}$Pb\tablenotemark{f}	&	5.93E-01	&	3.43E-01	&	10.6	&	&	& &	
1.84E-01	&	10.6	&	&	\nl
$^{207}$Pb\tablenotemark{f}	&	6.44E-01	&	4.10E-01	&	9.4	&	&	& &	
1.90E-01	&	9.4	&	&	\nl
$^{208}$Pb\tablenotemark{f}	&	1.83E+00	&	6.30E-01	&	8.4	&	&	& &	
1.81E-01	&	8.4	&	&	\nl
{\bf Pb}\tablenotemark{d}    &  & 46\%                          &           &                 &           & &       19\%  &                  
&                  &           \nl
\tablevspace{10pt}
$^{209}$Bi\tablenotemark{f}	&	1.44E-01	&	7.07E-03	&	13.1	&	1.37E-01	&	8.6	& &	
4.88E-03	&	13.1	&	1.39E-01	&	8.5	\nl
{\bf Bi}\tablenotemark{d}    & & 4.9\%                         &           &                 &           & &     3.4\% &                  
&                  &           \nl
\enddata
\tablenotetext{a}{The abundance distributions are normalized to $^{150}$Sm.}
\tablenotetext{b}{Anders, \& Grevesse (1989)}
\tablenotetext{c}{Between Cu and Sr the contribution of the weak $s$ component has been considered via the single exposure 
calculation of Beer, Walter, \& K\"appeler (1992). In this region the $\delta N_r$ 
uncertainties are omitted, since this modification is 
beyond the scope of the present paper.}
\tablenotetext{d}{The final line for each element denotes the contribution in percent of the main $s$ component to the solar 
elemental abundance.}
\tablenotetext{e}{Percent values in boldface denote an overabundance with respect to solar.}
\tablenotetext{f}{An important contribution from low metallicity stars is to be expected according to Gallino et al. (1998).}
\tablenotetext{g}{A contribution of $\sim6$\% from the weak $s$-component is to be expected (Raiteri et al. 1993).}
\end{deluxetable}

\newpage

\figcaption[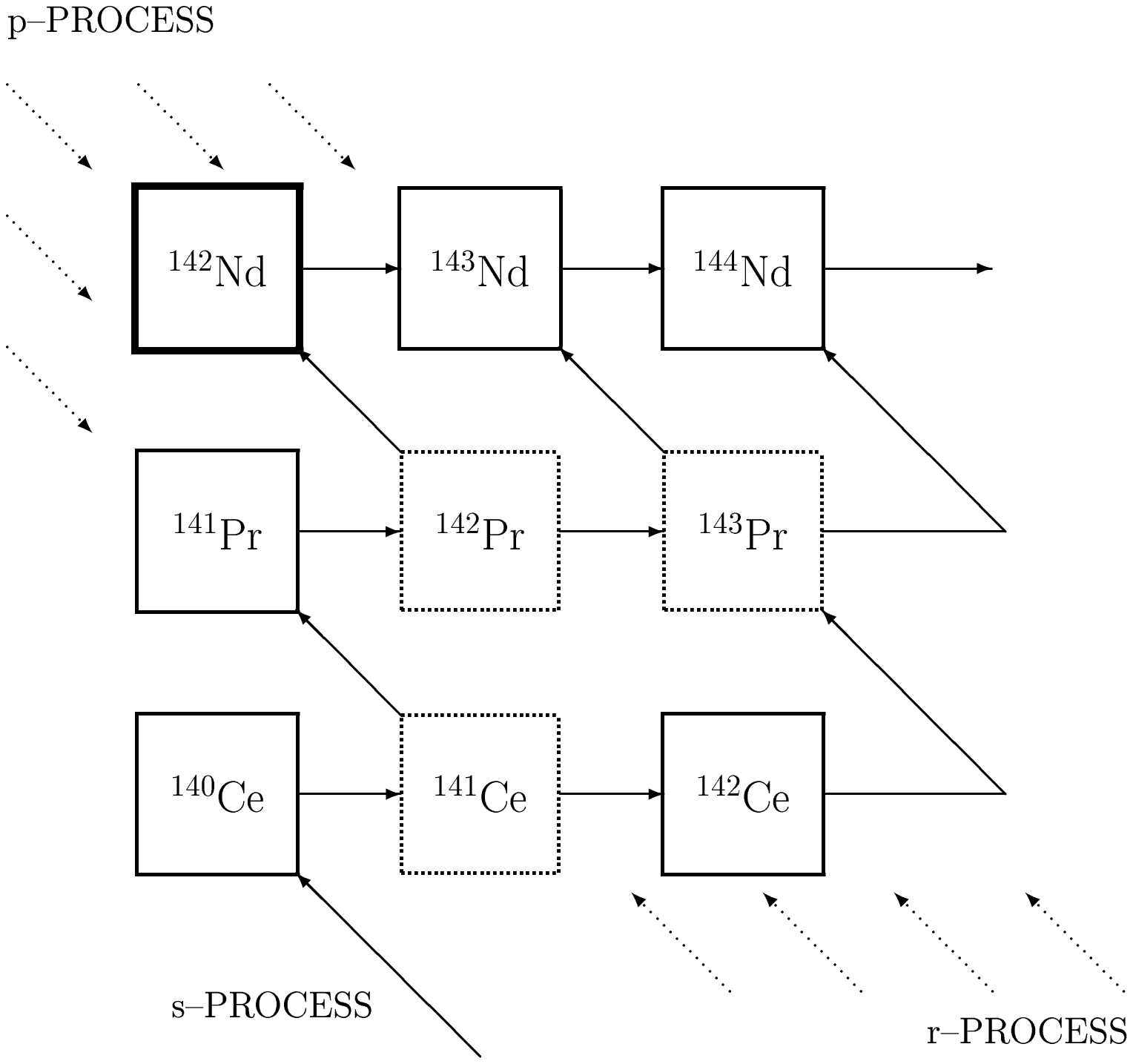]{The $s$-process reaction path in the Ce to Nd 
region, showing the branchings at $A$ = 141 and 142
that cause part of the reaction flow to partly bypass 
$^{142}$Nd. \label{fig1}}

\figcaption[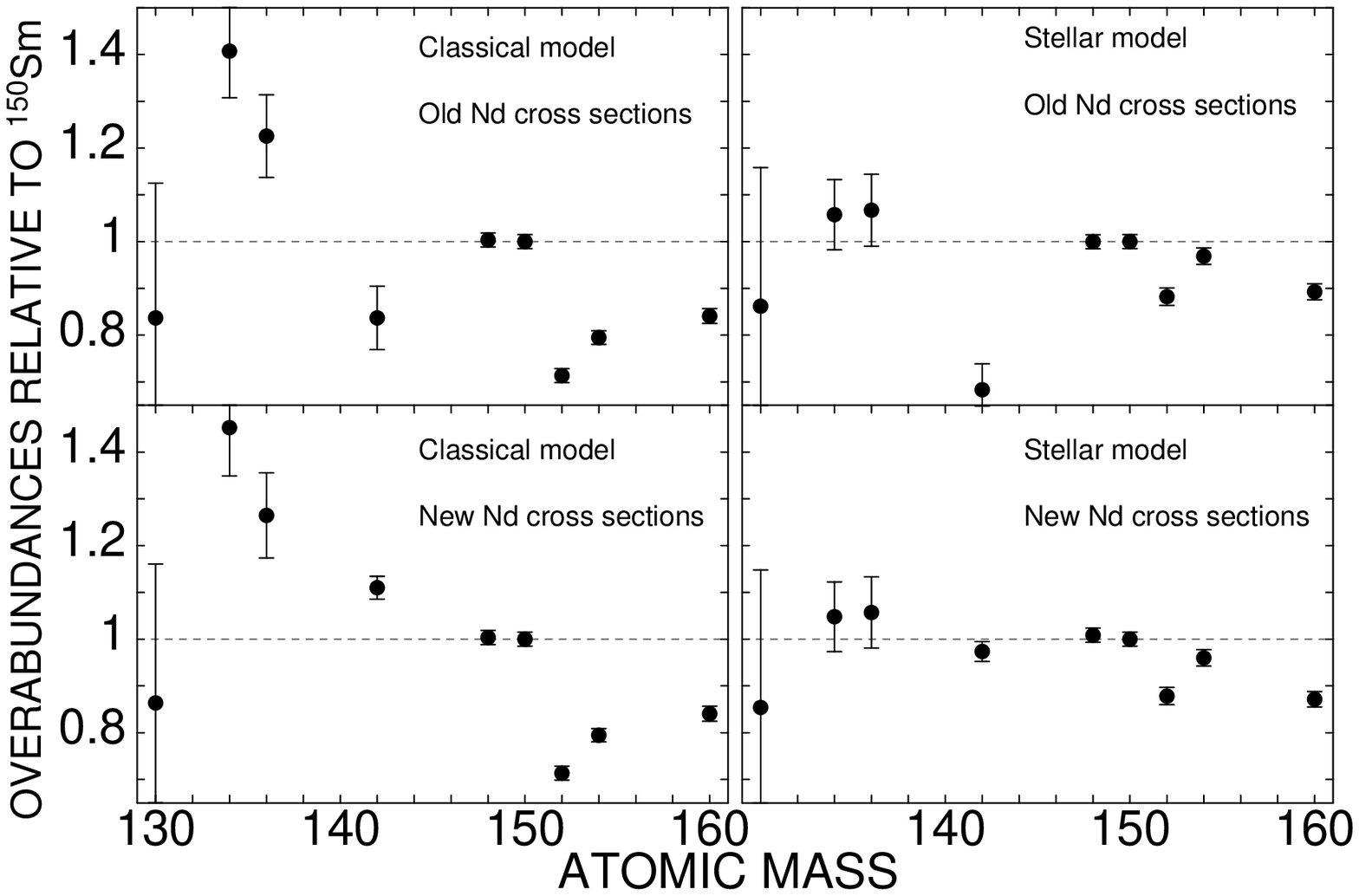]{The $s$-process overproduction in the 120 $<$ $A$ $<$ 160 region obtained with
the classical approach (left panels) and with the 2 $M_{\odot}$
"standard" model at $Z$ = 1/2 \zsb (right panels). The values are normalized to $^{150}$Sm.
The top panels represent the results
with the old Nd isotopes neutron capture cross sections, while the
effect of the new data is plotted in the bottom panels. The uncertainties on both solar abundances and cross sections are 
taken into account. \label{fig2}}

\figcaption[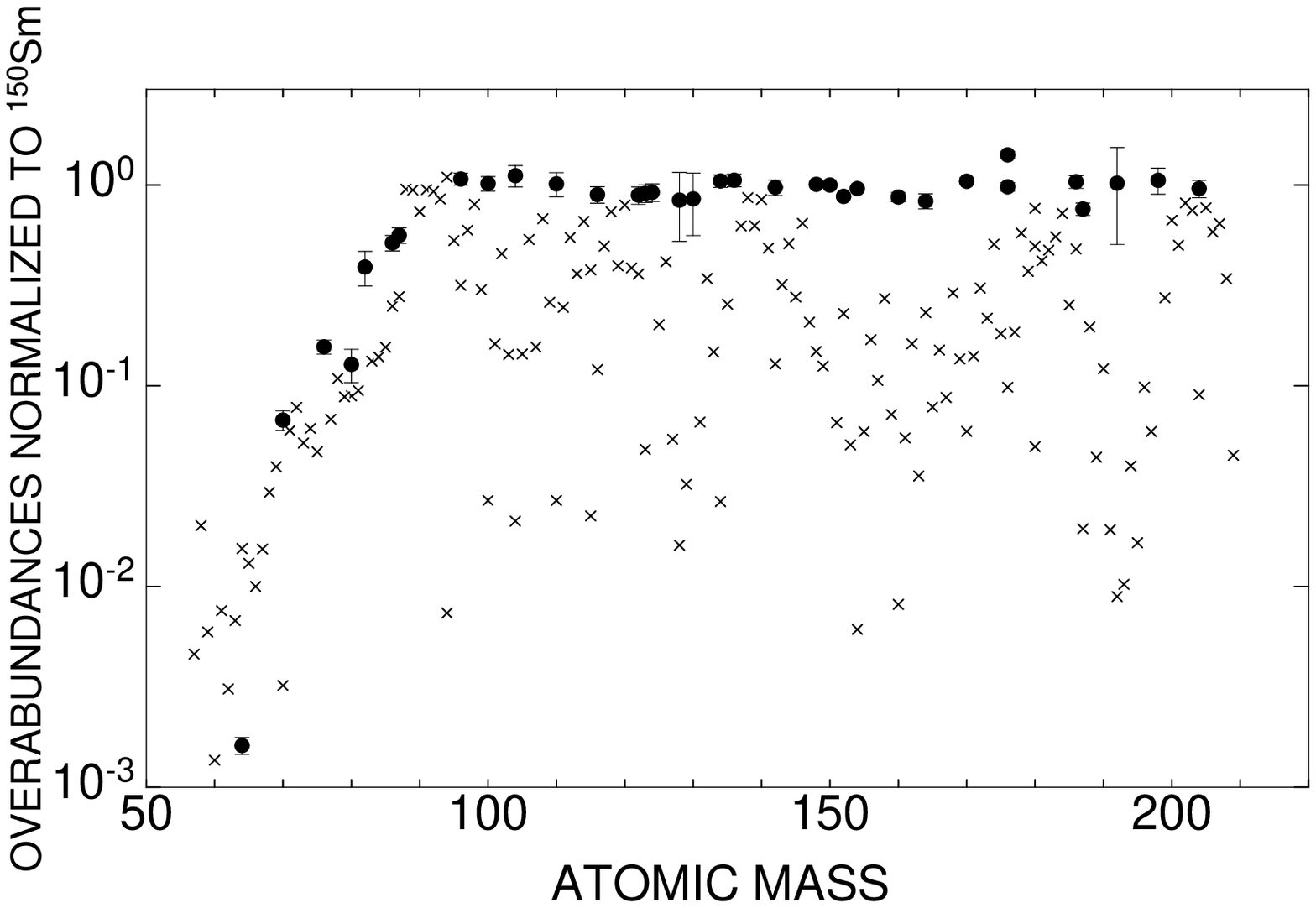]{The $s$-process abundance distribution that best reproduces 
the solar sistem main $s$ component,
as obtained by the stellar model for a 1.5 \msb of $Z$ = 1/2 \zsb 
("standard model") with updated Nd cross sections. The abundances are plotted as overproduction factors with 
respect to the solar values, normalized to $^{150}$Sm. Circles indicate $s$-only nuclei. The uncertainties on both solar 
abundances and cross sections are taken into account.  \label{fig3}}

\figcaption[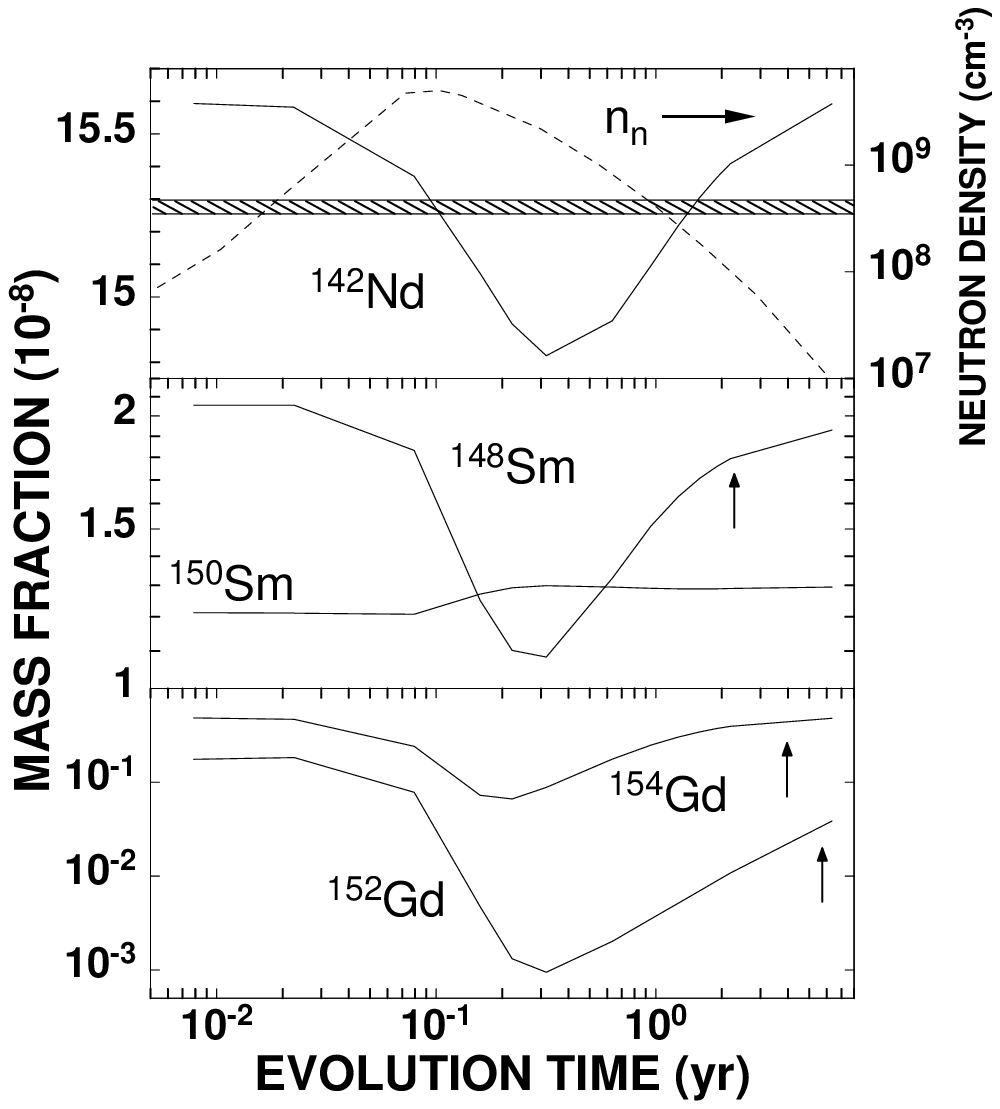]{The evolution of neutron density (top panel, right scale)
and of the abundances, plotted as fractions of mass, 
of some relevant isotopes during the \ndb neutron release, 
for a typical advanced pulse (pulse 15 of the standard AGB model). 
The time scale starts at the moment of bottom temperature reaching 2.5 $\times 10^8$ K. 
The effective neutron density, as derived from the classical approach, 
is sketched in the top panel by a shaded band. In the top panel is plotted 
the evolution of the \ndb abundance, while the same for $^{148}$Sm and 
$^{150}$Sm is found in the middle panel, as for $^{152}$Gd and $^{154}$Gd in the bottom panel. The arrows 
indicate the freeze-out instants, as indicate by the criterion $X_{freeze} = 0.9 X_{final}$. \label{fig4}}

\figcaption[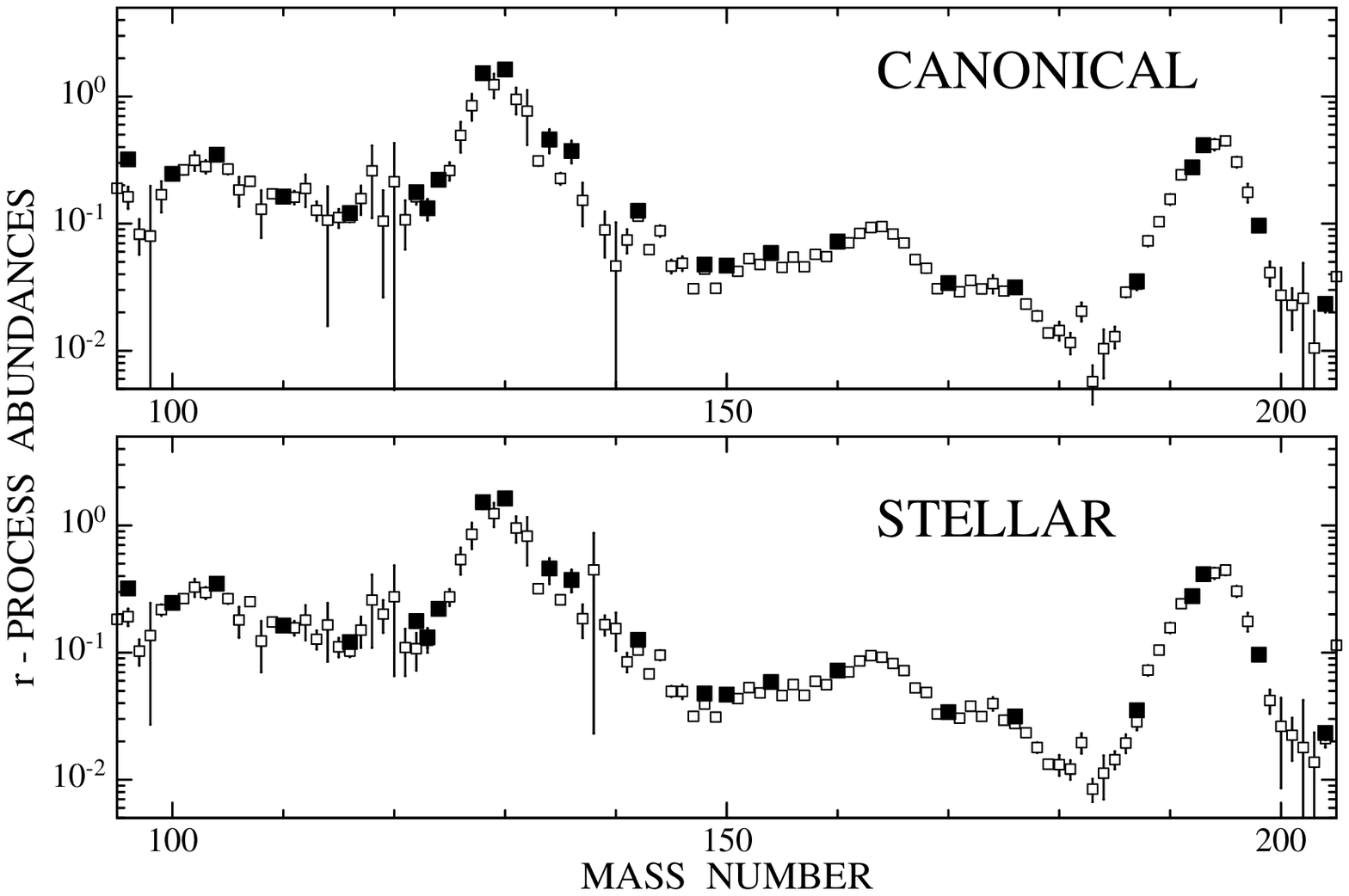]{The $r$-process residuals obtained by subtraction 
of the $s$-abundances obtained for the classical and for the stellar model.
In this case, as a somewhat closer representation of the Galactic chemical evolution 
mechanism, an average of the distributions best reproducing the main component 
for the standard 1.5 and 3 \msb models at $Z$ = 1/2 \zsb is considered. \label{fig5}}


\clearpage


\begin{thebibliography}{}

\bibitem[Anders \& Grevesse 1989]{ag89} Anders, E., \& Grevesse, N. 1989, \gca , 53, 197
\bibitem[Arlandini et al. 1996]{ap96} Arlandini, C., Gallino, R., Busso, M., \& Straniero, O. 1995, 
in Stellar Evolution: what should be done, ed. A. Noels, D. Fraipont-Caro, M. Gabriel, N. Grevesse, \& P. 
Demarque 
(Li\`ege: Univ. de Li\`ege), 447
\bibitem[Arlandini, K\"appeler, \& Wisshak 1998]{a98} Arlandini, C., K\"appeler, F., \& Wisshak, K. 1998,
Verhandl. DPG , 33, 456
\bibitem[Auble 1983]{au83} Auble, R. 1983, Nucl. Data Sheets, 40, 301
\bibitem[Bao \& K\"appeler 1987]{bk87} Bao, Z. Y., \& K\"appeler, F. 1987, Atom. Data Nucl. Data Tables, 36, 
411
\bibitem[Beer 1991]{b91} Beer, H. 1991, \apj, 379, 409
\bibitem[Beer, Corvi, \& Mutti 1997]{bp97} Beer, H., Corvi, F., \& Mutti, P. 1997, \apj , 474, 843
\bibitem[Beer \& K\"appeler 1980]{bk80} Beer, H., \& K\"appeler, F. 1980, \prc, 21, 534
\bibitem[Beer \& Macklin 1989]{bm89} Beer, H., \& Macklin, R. L. 1989, \apj, 339, 962
\bibitem[Beer et al. 1992]{bvw92} Beer, H., Voss, F., \& Winters, R. R. 1992, \apjs , 
80, 403 
\bibitem[Beer, Walter, \& K\"appeler 1992]{bwk92} Beer, H., Walter, G., \& K\"appeler, F. 1992, \apj, 389, 784
\bibitem[Best 1996]{b96} Best, J. 1996, Forschungszentrum Karlsruhe Scientific Reports, FZKA 5824 
(unpublished)
\bibitem[Busso et al. 1992]{bp92} Busso, M., Gallino, R., Lambert, D. L., Raiteri, C. M., \& Smith, 
V.V. 1992, \apj, 399, 218
\bibitem[Busso, Gallino, \& Wasserburg 1999]{bgw99} Busso, M., Gallino, R., \& Wasserburg, G. J. 1999,
\araa, in press
\bibitem[Busso et al. 1995]{bp95} Busso, M., Lambert, D. L., Gallino, R., Beglio, L., Raiteri, 
C. M., \& Smith, V. V. 1995, \apj, 446, 775
\bibitem[Busso et al. 1988]{bp88} Busso, M., Picchio, G., Gallino, R., \& Chieffi, A. 
1988, \apj, 326, 196
\bibitem[Busso et al. 1999]{bp99} Busso, M., Travaglio, C., Gallino, R., Lugaro, M., \& Arlandini, 
C. 1999, in Nuclei in the Cosmos V, ed. N. Prantzos, \& S. Harissopulos (Paris: Editions
Fronti\`eres), 227
\bibitem[Chieffi, Limongi, \& Straniero 1998]{cls98} Chieffi, A., Limongi, M., \& Straniero, O. 1998,
\apj, 478, 332
\bibitem[Chieffi, \& Straniero 1989]{cs89} Chieffi, A., \& Straniero, O. 1989, \apjs, 71, 47  
\bibitem[Clayton 1968]{c68} Clayton, D. D. 1968, Principles of Stellar Evolution and Nucleosynthesis 
(McGraw-Hill: New York)
\bibitem[Clayton et al. (1961)]{cp61} Clayton, D. D., Fowler, W. A., Hull, T. E., \& Zimmermann, B. A. 
1961, Ann. Phys., 12, 331
\bibitem[Clayton, \& Rassbach 1967]{cr67} Clayton, D. D., \& Rassbach, M. E. 1967, \apj, 148, 69  
\bibitem[Clayton \& Ward (1974)]{cw74} Clayton, D. D., \& Ward, R. A. 1974, \apj , 193, 397
\bibitem[Cosner, Iben, \& Truran 1980]{cit80} Cosner, K., Iben, I., Jr., \& Truran, J. W. 1980, 
\apjl, 238, L91
\bibitem[Cowan et al. 1980]{cp80} Cowan, J. J., Cameron, A. G. W., \& Truran, J. W. 1980, \apj , 241, 
1090 
\bibitem[De Laeter, Rosman, \& Ly 1998]{drl98} De Laeter, J. R., Rosman, K. J. R., \& Ly, C. 1998, 
Meteoritics \& Planet. Sci. Suppl., 33, A40
\bibitem[Denker et al. 1995]{dp95} Denker, A., Drotleff, H. W., Gro$\beta$e, M., Knee, H., Kunz, 
R., Mayer, A., Seidel, R., Soin\'e, M., W\"ohr, A., Wolf, G., \& Hammer, J. W. 1995, in
Nuclei in the Cosmos III, ed. M. Busso, R. Gallino, \& C. M. Raiteri (New York: American 
Institute of Physics Press), 255
\bibitem[Doll et al. 1999]{dp99}Doll, C., B\"orner, H. G., Jaag, S., Käppeler, F., \& Andrejtscheff W.
1999, \prc, 59, 492  
\bibitem[Frost \& Lattanzio 1996]{fl96} Frost, C. A., \& Lattanzio, J. L. 1996, \apj, 473, 383
\bibitem[Frost \& Lattanzio 1998]{fl98} Frost, C. A., \& Lattanzio, J. L. 1998, \apj, 500, 355
\bibitem[Gallino et al. 1998]{gp98} Gallino, R., Arlandini, C., Busso, M., Lugaro, M., Travaglio, 
C., Straniero, O., Chieffi, A., \& Limongi, M. 1998, \apj, 497, 388
\bibitem[Gallino et al. 1997]{gp97} Gallino, R., Busso, M., \& Lugaro, M. 1997, in 
Astrophysical Implications of the Laboratory Study of Presolar Materials, ed. T. Bernatowicz 
\& E. Zinner, (Woodbury: American Institute of Physics Press), 115
\bibitem[Gallino et al. 1999]{gp99} Gallino, R., Busso, M., Lugaro, M., Travaglio, C., Arlandini, C., \& 
Vaglio, P. 1999, in Nuclei in the Cosmos V, ed. N. Prantzos, \& S. Harissopulos 
(Paris: Editions Fronti\`eres), 216
\bibitem[Gallino et al. 1988]{gp88} Gallino, R., Busso, M., Picchio, G., Raiteri, C. M., \& 
Renzini, A. 1988, \apjl, 334, L45
\bibitem[Goriely (1997)]{g97} Goriely, S. 1997, \aap, 327, 845
\bibitem[Guber et al. 1997]{gup97} Guber, K. H., Spencer, R. R., Koehler, P. E., \& Winters, R. R. 
1997, \prl , 78, 2704
\bibitem[Harris (1981)]{h81} Harris, M. J.  1981, \apss, 77, 357
\bibitem[Hollowell \& Iben 1988]{hi88} Hollowell, D. E., \& Iben, I., Jr. 1988, \apjl , 333, L25
\bibitem[Holmes et al. (1976)]{hp76} Holmes, J. A., Woosley, S. E., Fowler, W. A., \& Zimmerman, B. A. 1976, 
Atom. Data Nucl. Data Tables, 18, 305
\bibitem[Howard 1991]{hpc91} Howard, W. M. 1991, private communication
\bibitem[Howard, Meyer, \& Woosley 1991]{hp91} Howard, W. M., Meyer, B. S., \& Woosley, S. E. 1991, 
\apjl, 373, L5
\bibitem[Jaag 1990]{j90} Jaag, S. 1990, Master Thesis, Univ. of Karlsruhe
\bibitem[Jaag \& K\"appeler 1996]{jk96} Jaag, S., \& K\"appeler, F. 1996, \apj , 464, 874
\bibitem[Jaag et al. 1999]{jp99} Jaag, S., Stoll, H., Wisshak, K., K\"appeler, F., Reffo, G.,
\& Rauscher, T. , \prc, in preparation
\bibitem[Jung et al. 1992]{jp92} Jung, M., et al. 1992, \prl, 69, 2164
\bibitem[K\"appeler, Beer, \& Wisshak 1989]{kap89} K\"appeler, F., Beer, H., \& Wisshak, K. 1989, 
Rept. Progr. Phys., 52, 945 
\bibitem[K\"appeler et al. 1982]{kap82} K\"appeler, F., Beer, H., Wisshak, K., Clayton, D. D., 
Macklin, R. L., \& Ward, R. A. 1982, \apj , 257, 821
\bibitem[K\"appeler et al. 1990]{kap90} K\"appeler, F., Gallino, R., Busso, M., Picchio, G., \& 
Raiteri, C.M. 1990, \apj, 354, 630
\bibitem[K\"appeler et al. 1996]{kap96} K\"appeler, F., Toukan, K. A., Schumann, M., \& Mengoni, 
A. 1996, \prc, 53, 1397
\bibitem[K\"appeler et al. 1994]{kap94} K\"appeler, F., Wiescher, M., Giesen, U., G\"orres, J., 
Baraffe, I., El Eid, M., Raiteri, C. M., Busso, M., Gallino, R., Limongi, M., \&  Chieffi, A.   
1994, \apj , 437, 396
\bibitem[Lambert 1995]{lpc95} Lambert, D. L. 1995, private communication
\bibitem[Lambert et al. 1995]{lp95} Lambert, D. L., Smith, V. V., Busso, M., Gallino, R., \& 
Straniero, O. 1995, \apj , 450, 302
\bibitem[Prantzos et al. 1990]{pp90} Prantzos, N., Hashimoto, M., Rayet, M., \& Arnould, M. 1990, 
\aap, 238, 455 
\bibitem[Raiteri et al. 1993]{rp93}Raiteri, C. M., Gallino, R., Busso, M., Neuberger, D., \& K\"appeler, F.
 1993, \apj, 419, 207 
\bibitem[Raiteri et al. 1999]{rp99}Raiteri, C. M., Villata, M., Gallino, R., Busso, M., \& Cravanzola, A. 1999,
\apjl, 518, L91 
\bibitem[Ratynski \& K\"appeler 1988]{rk88} Ratynski, W., \& K\"appeler, F. 1988, \prc, 37, 595
\bibitem[Rayet 1991]{rpc91} Rayet, M. 1991, private communication
\bibitem[Rayet et al. 1995]{rp95} Rayet, M., Arnould, M., Hashimoto, M., Prantzos, N., \& Nomoto, K.
1995, \aap, 298, 517
\bibitem[Rayet, Prantzos, \& Arnould 1990]{rp90} Rayet, M., Prantzos, N., \& Arnould, M. 1990, 
\aap, 227, 271 
\bibitem[Schumann et al. 1998]{sp98} Schumann, M., K\"appeler, F., B\"ottger, R., \& 
Sch\"olermann H. 1998, \prc, 58, 1790
\bibitem[Seeger, Fowler, \& Clayton (1965)]{sp65} Seeger, P. A., Fowler, W. A., \& Clayton, D. D. 1965, 
\apjs , 11, 121
\bibitem[Smith \& Lambert 1985]{sl85} Smith, V. V., \& Lambert, D. L. 1985, \apj , 294, 326
\bibitem[Smith \& Lambert 1986]{sl86} Smith, V. V., \& Lambert, D. L. 1986, \apj , 311, 843
\bibitem[Smith \& Lambert 1990]{sl90} Smith, V. V., \& Lambert, D. L. 1990, \apjs , 72, 387  
\bibitem[Straniero et al. 1997]{sp97} Straniero, O., Chieffi, A., Limongi, M., Busso, M., 
Gallino, R., \& Arlandini, C. 1997, \apj , 478, 332   
\bibitem[Straniero et al. 1995]{sp95} Straniero, O., Gallino, R., Busso, M., Chieffi, A., 
Raiteri, C. M., Salaris, M., \& Limongi, M. 1995, \apjl , 440, L85   
\bibitem[Takahashi \& Yokoi 1987]{ty87} Takahashi, K., \& Yokoi, K. 1987, Atom. Data Nucl. Data 
Tables, 36, 375 
\bibitem[Toukan et al. 1995]{tp95} Toukan, K. A., Debus, K., K\"appeler, F., \& Reffo, G. 1995, 
\prc , 51, 1540 
\bibitem[Travaglio et al. 1999]{tp99} Travaglio, C., Galli, D., Gallino, R., Busso, M., Ferrini, F., \&
Straniero, O. 1999, \apj, in press
\bibitem[Truran \& Iben 1977]{ti77} Truran, J. W., \& Iben, I., Jr. 1977, \apj , 216, 797
\bibitem[Ulrich 1973]{ul73} Ulrich, R. K. 1973, in Explosive Nucleosynthesis, ed. D. N. Schramm \& 
W. D. Arnett (Austin: University of Texas Press), 139
\bibitem[Vaglio et al. 1999]{vap99} Vaglio, P., Gallino, R., Busso, M., Travaglio, C., Straniero, O., 
Chieffi, A., Limongi, M., Lugaro, M., \& Arlandini, C. 1999, in Nuclei in the Cosmos V, 
ed. N. Prantzos, \& S. Harissopulos (Paris: Editions Fronti\`eres), 223
\bibitem[Voss et al. 1999]{vp99} Voss, F., Wisshak, K., Arlandini, C., K\"appeler, F., Kazakov, L., \& 
Rauscher, T. 
1999, \prc, 59, 1154
\bibitem[Voss et al. 1994]{vp94} Voss, F., Wisshak, K., Guber, K., K\"appeler, F., \& Reffo, G. 
1994, \prc, 50, 2582 
\bibitem[Ward, \& Newman 1978]{wn78} Ward, R. A., \& Newman, M. J. 1978, \apj, 219, 195  
\bibitem[Ward, Newman, \& Clayton 1976]{wp76} Ward, R. A., Newman, M. J., \& Clayton, D. D, 1976, 
\apjs , 31, 33 
\bibitem[Wisshak et al. 1990]{wip90} Wisshak, K., Guber, K., K\"appeler, F., Krisch, J., 
M\"uller, H., Rupp, G. \& Voss, F. 1990, Nucl. Instr. Meth. A, 292, 595 
\bibitem[Wisshak et al. 1993]{wip93} Wisshak, K., Guber, K., Voss, F., K\"appeler, F., \& Reffo, 
G. 1993, \prc, 48, 1401
\bibitem[Wisshak et al. 1998b]{wip98b} Wisshak, K., Voss, F., \& K\"appeler, F. 1998, \prc, 57, 3452
\bibitem[Wisshak et al. 1995]{wip95} Wisshak, K., Voss, F., K\"appeler, F., Guber, K., Kazakov, 
L., Kornilov, N., Uhl, M., \& Reffo, G. 1993, \prc, 52, 2762
\bibitem[Wisshak et al. 1998a]{wip98a} Wisshak, K., Voss, F., K\"appeler, F., Kazakov, L., \& 
Reffo, G. 1998, \prc, 57, 391
\bibitem[Wisshak et al. 1996]{wip96} Wisshak, K., Voss, F., Theis, C., K\"appeler, F., Guber, K., 
Kazakov, L., Kornilov, N., \& Reffo, G. 1996, \prc, 54, 1541 


\end{thebibliography}
\end{document}